\documentclass[authoryear,preprint,review,12pt]{elsarticle}

\usepackage{amssymb}
\usepackage{url}
\usepackage{siunitx}
\usepackage{amsmath}

\usepackage{lineno}
\journal{npj Urban Sustainability}

\begin{document}

\begin{frontmatter}



\title{Universal Patterns in the Long-term Growth of Urban Infrastructure in U.S. Cities from 1900 to 2015}



\author[1]{Keith Burghardt}
\ead{keithab@isi.edu}
\author[2,3,4]{Johannes H.Uhl}
\author[1]{Kristina Lerman}
\author[2,5]{Stefan Leyk}

\affiliation[1]{organization={USC Information Sciences Institute}, \addressline={4676 Admiralty Way}, city={Marina del Rey}, postcode={90292}, state={CA}, country={USA}}
\affiliation[2]{organization={Institute of Behavioral Science, University of Colorado Boulder}, address={483 UCB}, city={Boulder}, postcode={80309}, state={CO}, country={USA}}
\affiliation[3]{organization={Cooperative Institute for Research in Environmental Sciences (CIRES), University of Colorado Boulder}, address={216 UCB}, city={Boulder}, postcode={80309}, state={CO}, country={USA}}
\affiliation[4]{organization={Joint Research Centre, European Commission}, address={Via Enrico Fermi 2749}, city={Ispra}, postcode={21027}, state={VA}, country={Italy}}
\affiliation[5]{organization={Department of Geography, University of Colorado Boulder}, address={216 UCB}, city={Boulder}, postcode={80309}, state={CO}, country={USA}}

\begin{abstract}
Despite the rapid growth of cities in the past century, our quantitative, in-depth understanding of how cities grow remains limited due to a consistent lack of historical data. Thus, the scaling laws between a city's features and its population as they evolve over time, known as temporal city scaling, is under-explored, especially for time periods spanning multiple decades. In this paper, we leverage novel data sources  such as the Historical Settlement Data Compilation for the U.S. (HISDAC-US), and analyze the temporal scaling laws of developed area, building indoor area, building footprint area, and road length and other road network statistics for nearly all metropolitan areas in the U.S. from 1900 to 2015. We find that scaling exponents vary dramatically between cities as a function of their size and location. Three notable patterns emerge. First, scaling law exponents imply many, but not all, metropolitan areas are becoming less dense and indoor area per capita increases as cities grow, in contrast to expectations. Second, larger cities tend to have a smaller scaling exponent than smaller cities. Third, scaling exponents (and growth patterns) are similar between nearby cities. These results show a long-term trend that could harm urban sustainability as previously dense populations are rapidly spreading out into undeveloped land. Moreover, the regional similarity of long-term urban growth patterns implies that city evolution and sustainability patterns are more interconnected than prior research has suggested. These results help urban planners and scientists understand universal, long-term patterns of city growth across the US. 
\end{abstract}




\end{frontmatter}



 \section{Introduction}
 


Cities appear to follow common patterns that suggest simple mechanisms underly the complexity of urban growth~(\citenum{Zipf1949,Eeckhout2004,Jiang2011,Rozenfeld2008,Rozenfeld2011,Gabaix2004,Batty2006,Ribeiro2023}). 
For example, city properties vary consistently with their size, a phenomenon known as city scaling (\citenum{Bettencourt2007,Bettencourt2013,Burghardt2021_city}). 
While these analyses are usually conducted on cross-sectional data (i.e., across all cities at one point in time), recent works have explored longitudinal data (i.e., describing a single city across different points in time) to study temporal scaling laws of growing cities~(\citenum{Batty2006,Louf2014,Bettencourt2020_longvscross,Keuschnigg2019,Depersin2018}). This seemingly subtle difference in analytical design reveals unexpectedly non-universal features, with scaling laws that deviate from traditional theory (\citenum{Bettencourt2013}) and vary substantially between cities (\citenum{Bettencourt2020_longvscross}).

\begin{figure*}
    \centering
    \includegraphics[width=\linewidth]{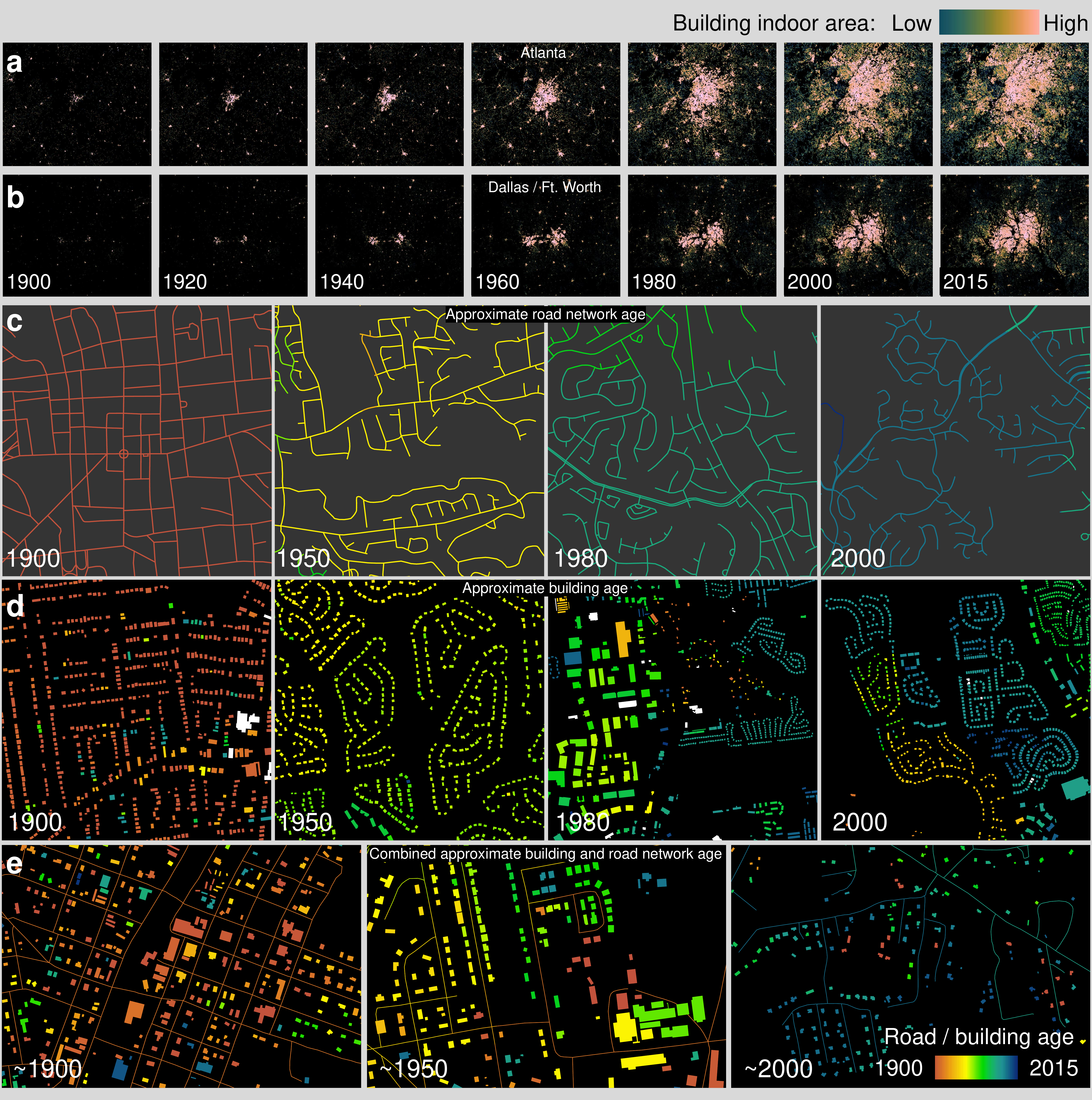}
    \caption{Data collected to analyze temporal scaling. (a) The growth of building indoor area over time for Atlanta, Georgia, and San Francisco, California. (b) Roads built at different ages (based on the ages of nearby houses). (c) Building footprints in different periods. Data from (\citenum{leyk2018hisdac,Burghardt2021_city}).
    }
    \label{fig:Schematic0}
\end{figure*}

In this paper, we explore these issues using vast, integrated geospatial datasets  to map urban infrastructure in over 850 metropolitan areas in the conterminous US (CONUS) at fine spatial resolution and over a broad time span. These datasets include the Historical Settlement Compilation for the U.S. (\citenum{leyk2018hisdac,Yoonjung2024}), Microsoft building footprint data (\citenum{MBF2020}), and road network data from the National Transportation Dataset (\citenum{NTD2020,Burghardt2021_city}), illustrated in Fig.~\ref{fig:Schematic0}. In contrast to previous analysis (\citenum{Samaniego2008,Strano2017,Lan2019,Bettencourt2007}), we are able to study temporal scaling over more than a century (1900-2015), constituting an observational window that goes beyond recent decades to which most geospatial data are constrained. Specifically, we analyze developed area, which encompasses the total built up land of a city, including their buildings, other built infrastructure and impervious surfaces, building indoor area, i.e., the total indoor residential area, commercial and industrial space, building footprint area, the total length of the urban road network, and a number of other road network statistics. 

Our results imply that, not only are smaller cities less dense than larger cities (\citenum{burghardt2023analyzing}), they are getting less dense more quickly over the course of urban expansion, which invades nearby undeveloped areas. Moreover, growing cities create ever larger buildings per person, which requires increasing energy for heating and cooling; this is an even greater problem in smaller cities. Finally, our analysis reveals strong correlations between the scaling laws of nearby cities which decrease very slowly with distance, meaning the urban development patterns of nearby cities are similar. 
Overall, these results offer new insights into the seemingly non-universal temporal patterns of city growth, with important implications regarding sustainability concerns in urban planning.


\section{Results}

\begin{table*}
    \centering
    \footnotesize
    \begin{tabular}{|p{4.2cm}|p{8.2cm}|}
 \hline
 \textbf{City statistic} & \textbf{Description}\\\hline\hline
 Developed area&  Total developed area in a city (\citenum{burghardt2023analyzing})  \\\hline
  Indoor area &Total indoor area in a city  in square meters  (\citenum{burghardt2023analyzing}) \\\hline
   Building footprint area &  Total area of building footprints in square meters  (\citenum{burghardt2023analyzing}) \\\hline
    Road length&  Total km of road built  in kilometers  (\citenum{burghardt2023analyzing}) \\\hline
    Number of deadends&  Number of cul-de-sacs and other ends of roads  (\citenum{Burghardt2021_city}) \\\hline  
    Number of 4+ intersections& Number of intersections where four or more edges (roads) meet  (\citenum{Burghardt2021_city}) \\\hline      
    Number of intersections&  Number of road intersections of all forms  (\citenum{Burghardt2021_city}) \\\hline   
    Number of edges&  Number of unique edges (defined to start at one intersection and end at another)  (\citenum{Burghardt2021_city}) \\\hline  
Population& Population in urban regions of a CBSA (\citenum{Burghardt2021_city}) \\\hline  
    \end{tabular}
    \caption{Urban statistics and population used to create temporal scaling.}
    \label{tab:descriptions}
\end{table*}
\subsection{Temporal scaling}
In contrast to cross-sectional analysis, which measures scaling relationships at a particular time across all cities, regardless of their age, temporal analysis tracks individual cities as they grow. One challenge here is defining city boundaries over time, because urban patches eventually coalesce as cities grow, making patch-level evolution difficult to assess. Because the largest one or two patches typically dominate the statistics (see SI Figure~S1), the definitions of city boundaries are less important for temporal analysis as the statistics ultimately tend to describe the largest city. Here instead, we define cities by their 2010 CBSA boundaries. We specifically analyze developed area, indoor area, building footprint area, road length, as well as the number of deadends, the number of intersections with four or more edges, the total number of intersections, and the total number of edges versus population (see definitions in Table~\ref{tab:descriptions}). All results are metric (area in meters squared or length in kilometers) or counts.

\begin{figure*}[tbh!]
    \centering
    \includegraphics[width=\linewidth]{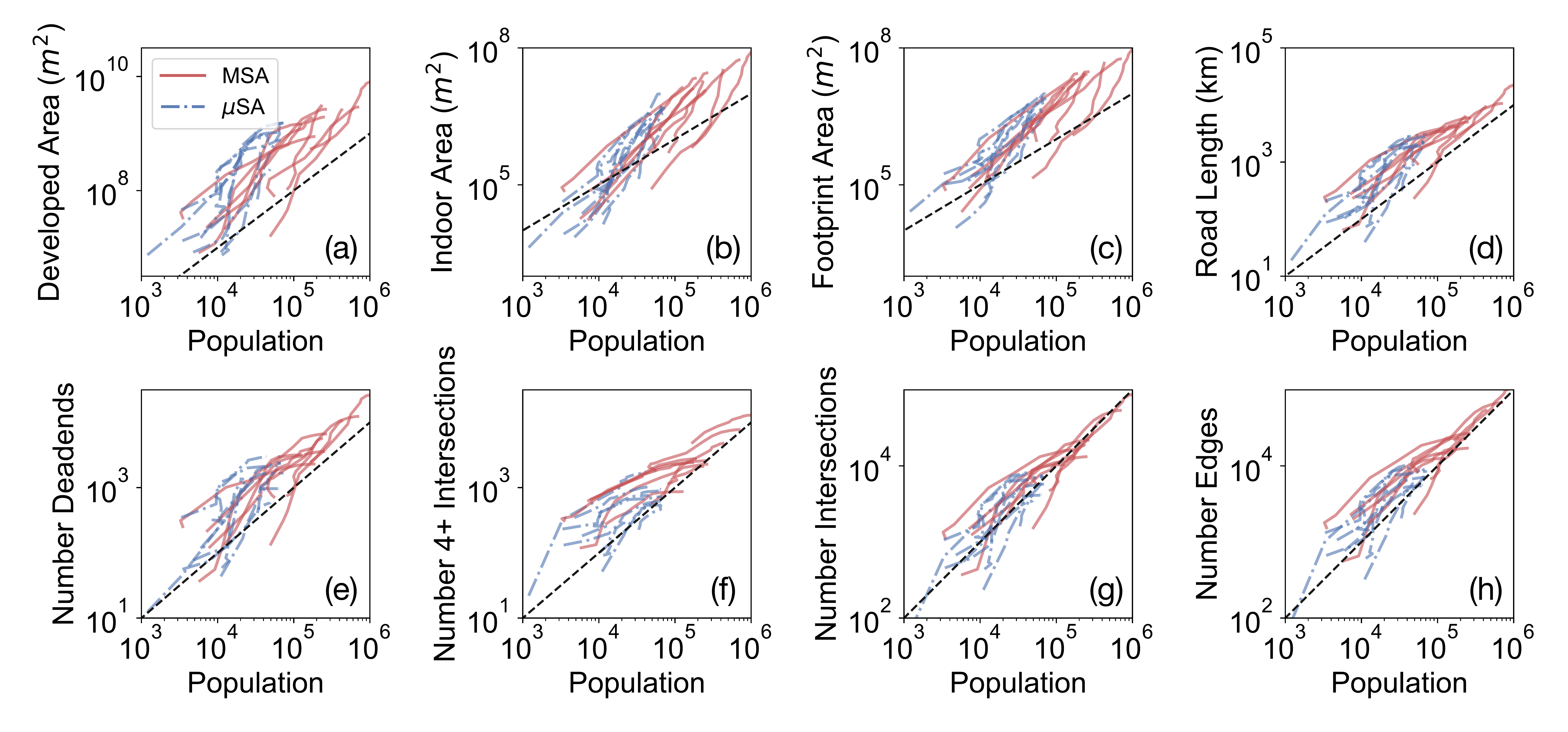}
    \caption{Temporal scaling for a random sample of cities. (a) Developed area, (b) indoor area, (c) building footprint area, (d) road length, (e) number of deadends, (f) number of intersections with four or more edges meeting, (g) total number of intersections, and (h) number of edges. The red lines represent 10 random MSAs (larger metropolitan areas) while blue dashed lines represent 10 random $\mu$SAs (micropolitan areas). Linear scaling is indicated by the black dashed line. }
    \label{fig:longitudinal_example}
\end{figure*}

We see very different scaling relations using temporal analysis. Figure~\ref{fig:longitudinal_example} shows network statistics as a function of growing population for a set of cities chosen at random from the data. 
We find scaling relations are often (but not always) superlinear for developed area, indoor area, building footprint area, and road length, with these statistics increasing faster than the population (linear scaling shown as a dashed line). These results imply that each new resident can be linked to increases in per-capita developed area, indoor area, building footprint area, and road length. In practice, this means that growing cities become less dense, and houses are getting larger. In contrast, the number of deadends, the number of road intersections with four or more edges, the total number of road intersections, and total number of edges often scales sub-linearly. These results suggest that as cities grow there are fewer new intersections and fewer edges of any length per capita. This is also consistent with a city becoming less dense.

\begin{figure*}
    \centering
    \includegraphics[width=0.8\linewidth]{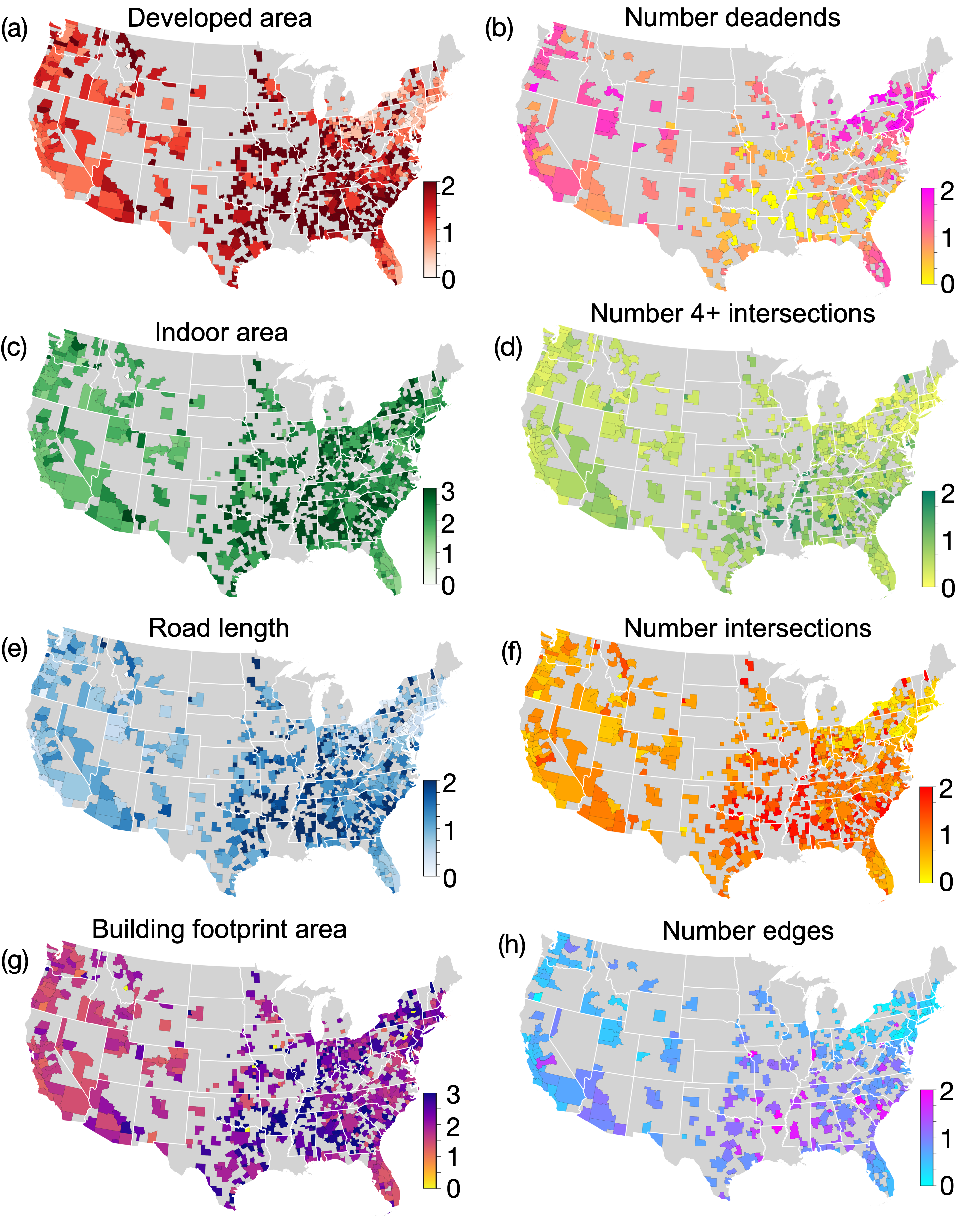}
\caption{Scaling laws across the US. Map of temporal scaling law exponents for (a) developed area, (b) number of deadends, (c) building footprint,  (d) number of 4+ intersections, (e) indoor area, (f) total number of intersections, (g) road length,  and (h) number of edges. Exponents greater than one indicate the statistic per capita increases as cities grow, while exponents less than one indicate a smaller statistic per capita as cities grow.}
    \label{fig:longitudinal_map}
\end{figure*}

\begin{figure*}
    \centering
    \includegraphics[width=0.8\linewidth]{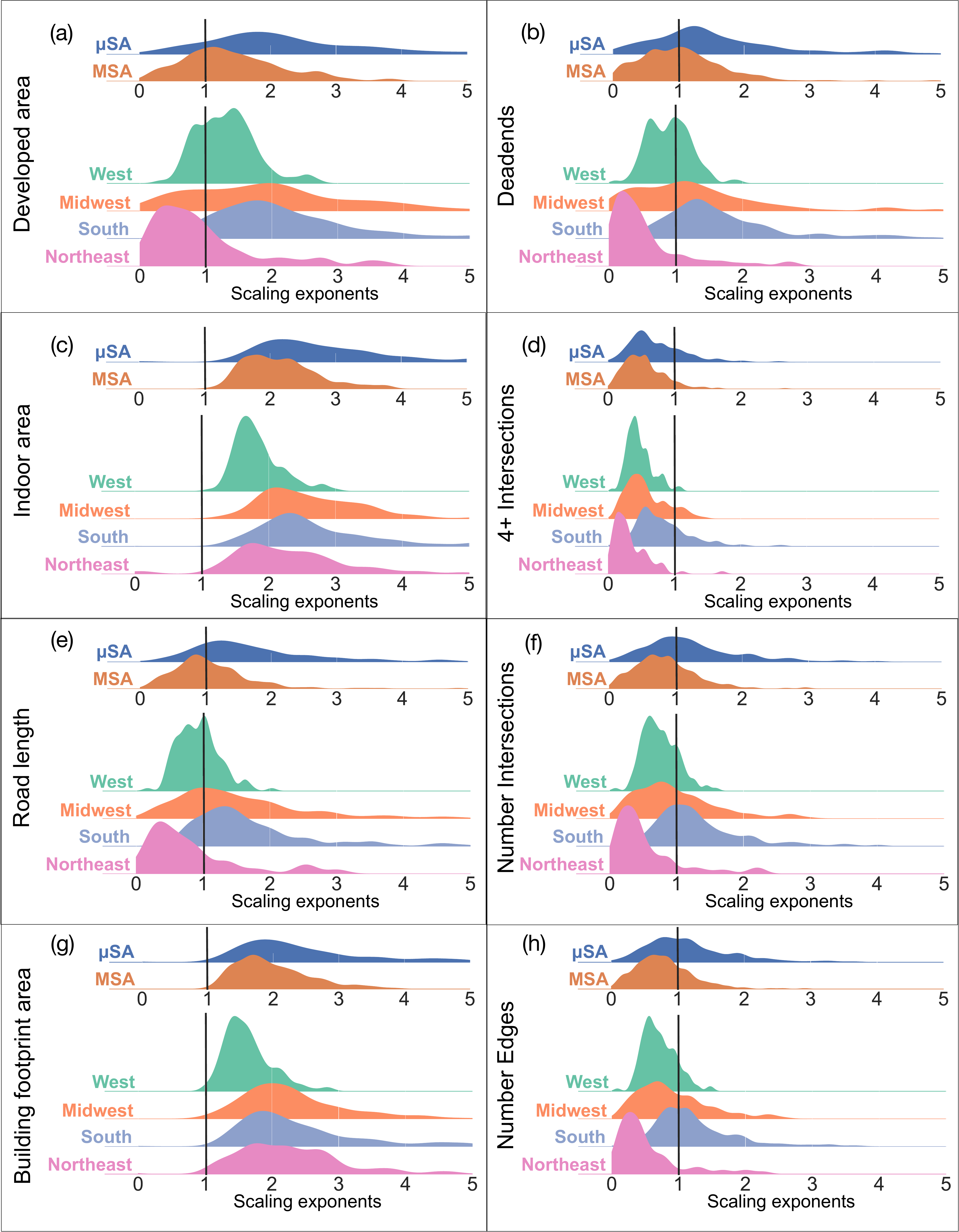}
    \caption{Scaling law exponents split by city size and region. Distribution of scaling exponents for (a) developed area, (b) number of deadends, (c) building footprint,  (d) number of 4+ intersections, (e) indoor area, (f) total number of intersections, (g) road length,  and (h) number of edges.
    Distributions are split by metro- (MSA) and micropolitan ($\mu$SA) statistical areas as well as US Office of Management and Budget-defined regions. Black vertical lines represent linear scaling. 
    }
    \label{fig:longitudinal_dist}
\end{figure*}

We see these results across many more cities in Figs.~\ref{fig:longitudinal_map} \& ~\ref{fig:longitudinal_dist}, which shows scaling laws for each of the statistics across the CONUS, as well as differences in these statistics for MSAs (larger cities, 317 in our analysis) and $\mu$SAs (smaller cities, 330 in our analysis) and across regions. We notably find that developed area and road length scale sublinearly for cities in the Northeast and West, but often scale \textit{superlinear} in the South and Midwest, with statistically significant differences between regions (see SI Figures~S2 \&~S3, based on a Dunns' posthoc test after Kruskal-Wallis test to compare distributions, p-value$<10^{-18}$ for each statistic). Meanwhile, indoor area and building footprint area mostly scale superlinearly, while road and intersection statistics almost always scale sublinearly. The wide distribution, however, means that some cities experience this effect more strongly than others. We notice scaling is most superlinear in $\mu$SAs, which are smaller micropolitan areas (with a population under 100,000 as of 2015). These areas include land surrounding Boone, NC; Athens, TX; or Findlay, OH.

While superlinear temporal scaling would seemingly disagree with theory (\citenum{Bettencourt2013}), it is consistent with other empirical analyses  (\citenum{Bettencourt2020_longvscross,Keuschnigg2019,Depersin2018}). These findings illustrate how, as cities grow, they achieve lower population densities (increasing developed area per person), a finding noticed outside of the city scaling literature (\citenum{Angel2017}). Why, however, is temporal scaling super-linear while cross-sectional scaling is often sub-linear (where the latter agrees with theory (\citenum{Bettencourt2013,Lobo2020}))? We can square this circle through prior cross-sectional analysis of these data  (\citenum{burghardt2023analyzing}), which show that, while larger cities are indeed denser than smaller cities, all cities become less dense as they grow. This is in agreement with US Census data, shown in Supplementary Figure S16 of (\citenum{burghardt2023analyzing}), where the average size of new houses, for example, has consistently increased over the past forty years. While cities both large and small have similar house sizes at a given point in time, there is a general evolution towards more spacious houses, possibly due to processes related to urban sprawl (\citenum{boeing2020off,barrington2015century,barrington2020global,Burghardt2021_city}). Similar results for developed area and road length may be due to the increasing use of cars and other long-range transport in the US. Namely, urban sprawl has likely also led to more kilometers of road covering thinly-populated suburbs, which then results in superlinear temporal scaling between road lengths and city population.


\begin{figure*}
    \centering
    \includegraphics[width=\linewidth]{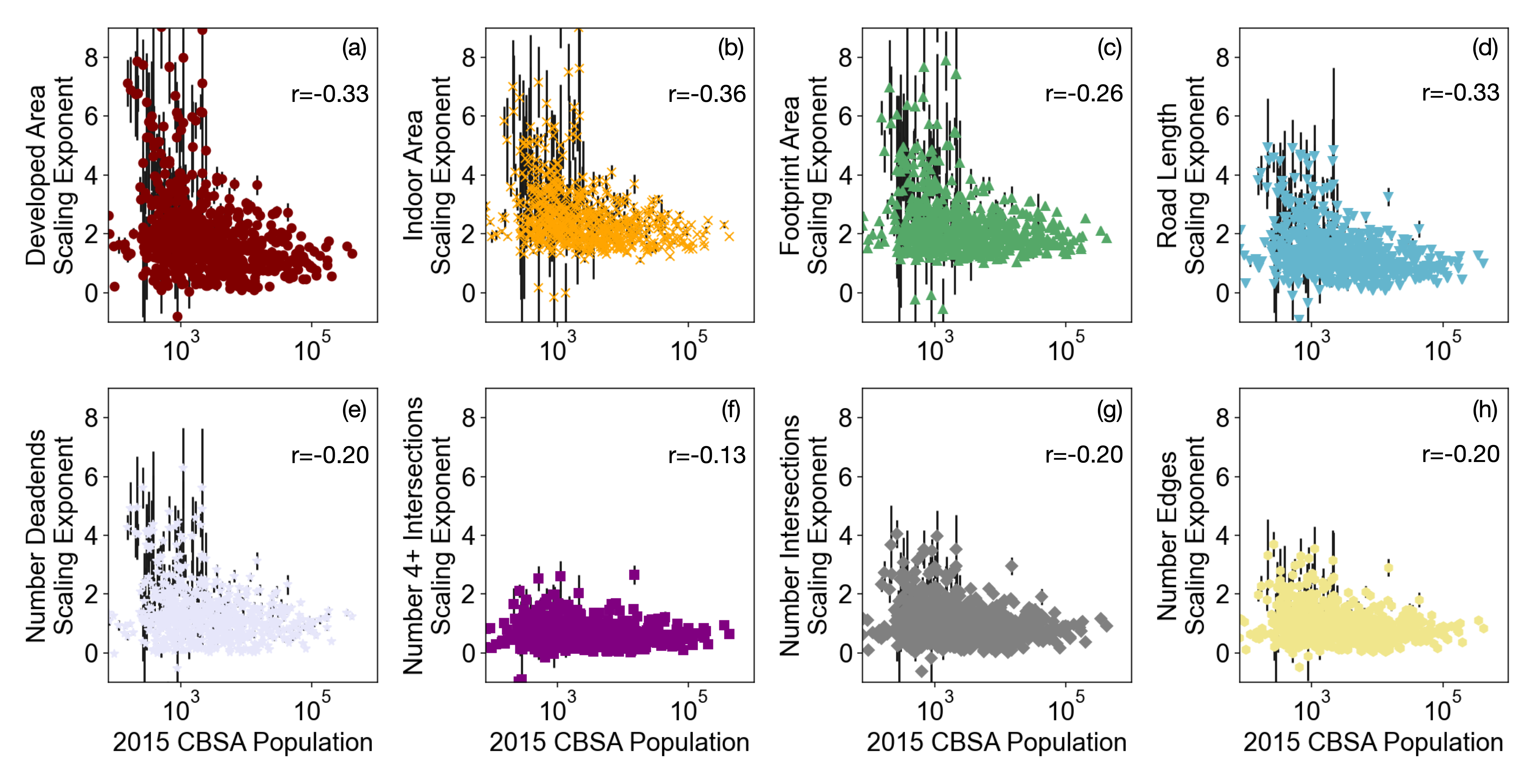}
    \caption{Scaling law exponent versus 2015 population. (a) Developed area, (b) indoor area, (c) building footprint area, (d) road length, (e) number of deadends, (f) number of 4+ intersections, (g) total number of intersections, and (h) number of edges. Black error bars represent standard errors of scaling law coefficients. Spearman correlations are shown in each figure (all correlations are statistically significant, p-values $<10^{-6}$).}
    \label{fig:scalingvspop}
\end{figure*}

\begin{figure*}
    \centering
    \includegraphics[width=0.8\linewidth]{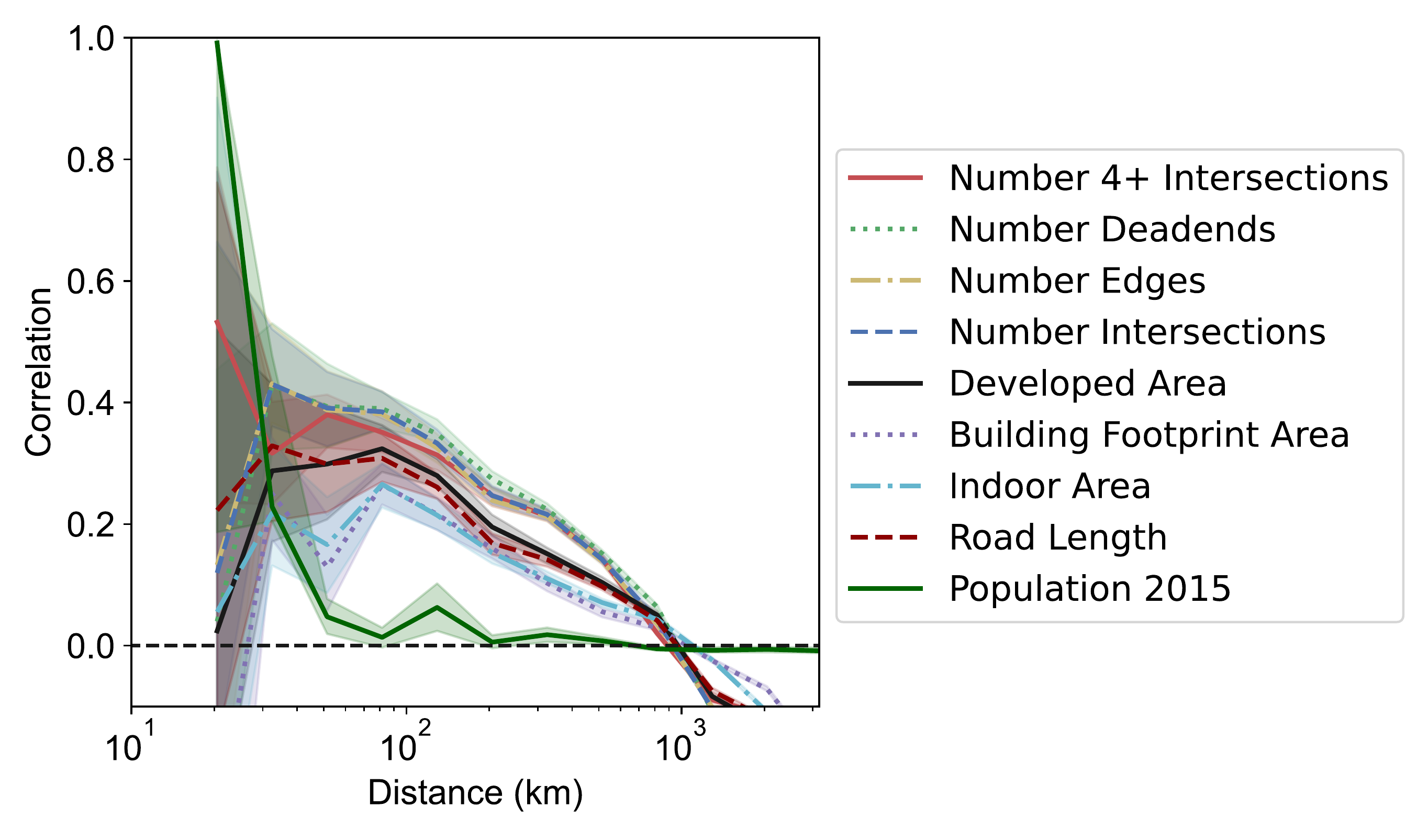}
    \caption{Correlations of scaling exponents versus distance between nearby cities. This figure shows the correlation of temporal scaling exponents versus distance for developed area, indoor area, building footprint area, road length, number of deadends, number of intersections with four or more edges meeting, total number of intersections, and number of edges.  Correlations versus distance for CBSA population is also shown. Shaded regions indicate 68\% confidence intervals in the mean correlation.}
    \label{fig:scaling_distance_correl}
\end{figure*}

\subsection{Spatial correlation}

We gain further knowledge of temporal scaling laws by analyzing Fig.~\ref{fig:longitudinal_map}). We notice that high or low scaling law exponents appear to cluster in close proximity to each other. We plot correlations of scaling law exponents over distance in Fig.~\ref{fig:scaling_distance_correl} which is inspired by plots presented by (\citenum{Gracia2014}) with respect to voting patterns. We expect other methods, such as variograms (\citenum{Deutsch2001encyclopedia}), will show similar results. As we might intuit, the closest cities have the most similar scaling law exponents. This correlation drops only as the log of distance until it reaches zero at about 1000km. We notice, however, that scaling laws correlate negatively with 2015 population, as seen in Fig.~\ref{fig:scalingvspop}. We therefore check if these long-range scaling correlations are trivially due to spatial correlations in population. However, we see that population correlations as of 2015 drop to zero after 100km, a range at which exponent spatial correlations still remain high. When we separate our analysis by MSA and $\mu$SA, apply different data filters, and apply a different city growth definition, the results broadly hold (SI Figures~S4,~S5,~S6,~S7,~S8,~S9,~S10,~S11,~S12, \&~S13). We therefore conclude that city population correlations cannot fully explain these results and other factors, such as topography or social differences may play a role as well. Interestingly, the correlation decreases more slowly for MSAs than $\mu$SAs (SI Figures~S6 \&~S10) therefore this long-range correlation appears strongest for large urban areas.

Interestingly, Fig.~\ref{fig:scalingvspop} also implies that larger cities have a lower scaling law compared to smaller cities. Therefore smaller cities become even less dense as they grow relative to larger cities. Larger cities, therefore exhibit a relative economy of scale that compounds over time. When we break up scaling analysis between time periods 1900--1950 and 1960--2015, Appendix~S11, we find that this trend is only found for scaling since 1960, which means early cities grew independent of city size. Only recently have smaller cities grown at much faster rates than larger cities.

\section{Discussion}

In this study, we have explored temporal scaling relations between city infrastructure and population and how it varies across space among 857 metropolitan areas in the CONUS since 1900. These results show significant variation in scaling laws, although they are often super-linear, meaning cities become less dense as they grow (\citenum{Angel2017}), and houses are getting larger (\citenum{burghardt2023analyzing}). This in turn means more area is being developed per person. Larger houses might imply more energy is being spent on materials and climate control.

Interestingly, we see long-range correlations in scaling laws, where nearby cities have similar scaling laws, and this correlation only drops slowly with distance. This is important because theory behind the temporal scaling laws of cities is lacking, but while it differs from cross-sectional scaling (i.e., scaling across cities at a given point in time (\citenum{Bettencourt2020_longvscross, burghardt2023analyzing}), there are still seemingly universal patterns that we can utilize to better understand this type of scaling. Moreover, the correlations are not a product of how scaling laws are calculated as we see similar results for other growth patterns of cities. The reasons for this behavior are numerous, but we will focus on three (non-exclusive) hypotheses:
\begin{enumerate}
    \item \textbf{Friction of distance.} A key principle of urban geography is that movement is costly (\citenum{gregory2011dictionary}). Aspects of city planning may, therefore, be sequestered by distance-based relationships. Urban planners of some areas may therefore have advocated for e.g., single family homes or building height regulations, while others may have advocated for more multi-family housing or more high-rises due to local regulations or schools of thought. This can have a direct effect on how many roads or houses are build and thus how much developed area, indoor area, building footprint area, or kilometers of road get built.
    \item \textbf{Topography.} Roads and their construction are strongly affected by topography, such as gradient (\citenum{Boeing2020multi,Burghardt2021_city}), and we expect this will be similar in nearby cities. For example, areas bordering waters, such as San Francisco, California and Portland, Oregon, will not expand outwardly as much as areas in a plain, such as Houston and Dallas, Texas. Moreover, regulations such as building codes for earthquake-proofing a building will itself affect how a city grows (e.g., increasing the expense of tall buildings to ensure they are earthquake-resistant). This in turn affects how population drives super- or sub-linear scaling with developed area, and therefore road lengths and indoor area between nearby cities. Far away cities, however, have less topography in common and therefore can grow in different ways.
    \item \textbf{Climate.} Finally, we find that climate is fairly localized (\citenum{Hu2018}), and thus may play a role in how cities develop. Cities that experience hurricanes may, for example, be built, and expand differently, than dryer locations. Miami, Florida, for example, enforces strong building codes to protect structures from hurricanes (\citenum{Done2018}), which in turn affect what buildings can be built (and possibly affect building height to ensure resistance to high winds). 
\end{enumerate}

Road network statistics, however, mildly buck the trends we see of other statistics. Namely, the correlations, especially of network degree, remain high even for cities that are far away. This may be due to broader trends in city planning, such as the tendency to move from grid-like planned cities to more irregular suburban road networks since cars became commonplace (\citenum{Burghardt2021_city}). That said, some city statistics are strongly distance dependent, such as orientation entropy, which measures the variability of road orientations.

We also notice that scaling laws are smaller in MSAs than $\mu$SAs, and there is a negative correlation between scaling laws and city population as of 2015. These results suggest that larger cities \emph{have a compounding economy of scale} relative to smaller cities, and therefore take up less additional land, even as they, like most other cities, get less dense over time. This adds to recent literature on the economy of scale in larger cities (\citenum{Bettencourt2007,burghardt2023analyzing}; larger cities are more compact at any given time, they also use their space more efficiently than smaller cities as they grow.

\subsection{Implications}
These results have important implications for sustainability, urban planning, and complex systems. First, lower-density urbanization is of grave concern for sustainability, as more land is being developed for fewer people, which could harm nearby ecosystems. This result is especially concerning for smaller cities, whose scaling laws are even higher, meaning their growth has a more notable impact on undeveloped land. In addition, increasing house sizes, especially in smaller cities, imply more materials are being used for each dwelling, and more energy might be used to heat and cool these structures.

Our results are informative to urban planners as schools of thought may shape the development of cities. Alternatively, these patterns may be a function of climate or topography, therefore methods to reduce urban sprawl should focus on topographic or climatic regions that enable such sprawl to happen rather than treating the problem the same way across all regions. Finally, this work is of importance to scientists studying complex systems as they imply the seemingly complex growth of cities is driven by simple fundamental patterns, such as scaling laws that are consistently super-linear and depend on relative city size and location. These results also imply city growth is strongly spatially correlated, which prior theoretical work has not yet explored.

\subsection{Limitations and Future Work}
There are certain limitations of the data we analyze. For example, building years are unknown for a number of records. Furthermore, we assume that roads are constructed at approximately the same time as nearby buildings and that population is proportional to the number of buildings within a given patch. We evaluated the reliability of such assumptions to the best of our ability, but more research is needed to test and improve upon them. For example, it is important to uncover new ways to approximate patch-level population estimations, as well as city infrastructure over time. Moreover, an ongoing issue is how to work with missing data. We have shown in previous work that many of our conclusions about city infrastructure are robust to missing data (\citenum{Burghardt2021_city}), but these results need to be explored in more detail, including uncovering ways to impute missing build-years or missing buildings within our data. Finally, there are many future ways to test the robustness of these results including determining temporal scaling laws in other datasets and measuring temporal scaling laws for alternative city boundary definitions (\citenum{burghardt2023analyzing}). 
\section{Conclusions}

We utilized a recent dataset to analyze 857 CONUS metropolitan areas longitudinally since 1900 via temporal scaling. We discover that temporal scaling diverges markedly from previous theory (\citenum{Bettencourt2013}) because cities have superlinear scaling, meaning as they grow they become less dense (\citenum{Angel2011,Angel2017}) and indoor area per capita is increasing (\citenum{burghardt2023analyzing}). We also find that nearby cities grow similarly, with a correlation that decreases slowly with distance. Moreover, larger cities have a lower scaling law exponent, meaning that while they become less-dense, they are relatively denser than smaller cities. These results appear robust despite variations in data quality across the US.

\section{Methods}

We recently developed a novel approach to capture the long term evolution of road networks for 857 CBSAs within the US since 1900~(\citenum{Burghardt2021_city}), and applied it to understand cross-sectional scaling (\citenum{burghardt2023analyzing}). This approach has a number of limitations, notably that some counties have better temporal completeness (have buildings with known dates) and spatial coverage (houses have spatial identifiers) than others. As in previous work (\citenum{burghardt2023analyzing,Burghardt2021_city}), we filter data to have temporal completeness $> 60\%$ and spatial coverage $> 40\%$. 647 CBSAs were considered complete enough for our analysis. In the SI, we also show data for temporal completeness and spatial coverage $>0\%$ (all 857 CBSAs) as well as $>80\%$ (64 CBSAs). All results are broadly consistent. 

When analyzing temporal scaling laws, we use CBSA boundaries as of 2010. We find in SI Figure S1 that these analyses are dominated by the largest city in a given CBSA, represented as a patch of developed area. This means that, while other cities grow in this metropolitan area, their contributions are minimal. Meanwhile, because patches merge together as they grow, it is difficult to otherwise find an uncontroversial objective definition of a city over time unless we take into account all patches that will eventually make up the city's metropolitan area. Alternative definitions of city boundaries, such as CBSA definitions from before 2010 do not provide a consistent metropolitan area. Namely, what counts as a CBSA has evolved as have their boundaries. Changes in CBSA definitions mean a smaller city may suddenly become part of a metropolitan area thereby skewing the observed scaling law. Moreover, $\mu$SA definitions did not exist before 2003 (\citenum{BLM_usa}) and MSAs are not defined before 1949 (\citenum{BLM_msa}).

Using these previous data, (\citenum{burghardt2023analyzing}), we aggregate populations across patches within a CBSA as well as various statistics, such as developed area, indoor area, and road length. We then track these data over time, as seen in Fig.~\ref{fig:longitudinal_example} and find the scaling coefficient as the best-fit of the log-scaled data for each city. We avoid non-linear fits as these are found to be highly sensitive to outliers (\cite{burghardt2023analyzing}). These results are plotted in Fig.~\ref{fig:longitudinal_map} and found to be robust to data filtering methods, as shown in SI Figures~S14, ~S15,~S16, \&~S17. These data closely fit a power-law, as shown in SI Figures~S18 \&~S19
. Moreover, the scaling results do not change dramatically if we split data $\le 1950$ or $>1950$, as seen in SI Figure~S20.

To calculate spatial correlations, we used the Python libraries \texttt{shapely} (\url{https://shapely.readthedocs.io/en/stable/}) and \texttt{geopy} (\url{https://geopy.readthedocs.io/en/stable/}) to determine CBSA centroids (namely centers of built-up patches as of 2015), and distances between centroids, respectively. We then use the Python library \texttt{scipy} (\url{https://scipy.org/}) to determine Pearson correlations within binned distance intervals. 

We explore infrastructure within high build intensity regions as cities grow. Our work has revealed strongly linear correlations between the number of houses in a city metropolitan area and its population. We use this finding to estimate population within each urban patch as the fraction of total houses existing within a particular CBSA multiplied by the CBSA's US census population. This simple assumption is in strong agreement with other metrics, including state-of-the-art population dasymetric refinement methods (see Supplementary material in (\citenum{burghardt2023analyzing})).

While extrinsic variables (developed area, building footprint area, indoor area, and road length) are all available from previous scaling analysis (\citenum{burghardt2023analyzing}), we calculate road network statistics for each decade using data from (\citenum{Burghardt2021_city}). We removed parts of the network visible in the previous decade to find parts of the road network seen in the current decade. From these networks, we calculated road density as the km of new road divided by the total amount of new developed area in square km. We also calculated the mean road intersections, fraction of 4+ road intersections, fraction of road deadends, and mean local griddedness using Python library \texttt{networkx} as these features only depend on the road network topology. Finally, orientation entropy was calculated as it was in (\citenum{Burghardt2021_city}) for new roads except we calculated the entropy of edge orientation angles, discretized into bins of $\ang{5}$. We summarize all our statistics and population definitions in Table~\ref{tab:descriptions}, based on (\citenum{Burghardt2021_city}).

Code and data for our analysis can be found at: \url{https://anonymous.4open.science/r/longitudinal_scaling-2004/README.md}.

\section{Data Availability}
The data is available in the following references: US regions (\citenum{CensusRegions}), population (\citenum{census_county_pop_old,census_county_pop_mid,census_county_pop_new}), built-up Property Records (\citenum{databupr2020}), built-up property locations (\citenum{databupl2020}), built-up area (\citenum{databua2020}), building footprint area (\citenum{BUFA}), and indoor area (\citenum{data_bui2018}), road networks (\citenum{Burghardt2022_roadstats}), and patch boundaries (\citenum{Uhl2022_urbanarea}). In these data, references to patches are density-based settlements within a CBSA. Merging any neighboring cross-CBSA patches (using (\citenum{Uhl2022_urbanarea}) creates the city features we analyze in this paper.

\section{Competing Interests}
The Authors declare no competing financial or non-financial interests.

\section{Author Contributions}
K.B., J.U., K.L., and S.L. designed research; K.B. performed analysis; K.B., J.U., K.L., and S.L. wrote the paper.

\section*{Acknowledgements}
Partial funding for this work was provided through the National Science Foundation (award number 2121976 ) and the Eunice Kennedy Shriver National Institute of Child Health and Human Development of the National Institutes of Health under award number P2CHD066613. The content is solely the responsibility of the authors and does not necessarily represent the official views of the National Institutes of Health. We acknowledge access to the Zillow Transaction and Assessment Dataset (ZTRAX) through a data use agreement between the University of Colorado Boulder and Zillow Group, Inc. More information on accessing the data can be found at http://www.zillow.com/ztrax. The results and opinions are those of the authors and do not reflect the position of Zillow Group. Moreover, Safe Software, Inc., is acknowledged for providing a Feature Manipulation Engine (FME) Desktop license used for data processing.


\begin{thebibliography}{52}
\expandafter\ifx\csname natexlab\endcsname\relax\def\natexlab#1{#1}\fi
\providecommand{\url}[1]{\texttt{#1}}
\providecommand{\href}[2]{#2}
\providecommand{\path}[1]{#1}
\providecommand{\DOIprefix}{doi:}
\providecommand{\ArXivprefix}{arXiv:}
\providecommand{\URLprefix}{URL: }
\providecommand{\Pubmedprefix}{pmid:}
\providecommand{\doi}[1]{\href{http://dx.doi.org/#1}{\path{#1}}}
\providecommand{\Pubmed}[1]{\href{pmid:#1}{\path{#1}}}
\providecommand{\bibinfo}[2]{#2}
\ifx\xfnm\relax \def\xfnm[#1]{\unskip,\space#1}\fi
\bibitem[{Ahn et~al.(2024)Ahn, Leyk, Uhl and McShane}]{Yoonjung2024}
\bibinfo{author}{Ahn, Y.}, \bibinfo{author}{Leyk, S.}, \bibinfo{author}{Uhl,
  J.H.}, \bibinfo{author}{McShane, C.M.}, \bibinfo{year}{2024}.
\newblock \bibinfo{title}{An integrated multi-source dataset for measuring
  settlement evolution in the united states from 1810 to 2020}.
\newblock \bibinfo{journal}{Scientific Data} \bibinfo{volume}{11},
  \bibinfo{pages}{275}.
\newblock \URLprefix \url{https://doi.org/10.1038/s41597-024-03081-x},
  \DOIprefix\doi{10.1038/s41597-024-03081-x}.
\bibitem[{Angel et~al.(2017)Angel, Parent, Civco and Blei}]{Angel2017}
\bibinfo{author}{Angel, S.}, \bibinfo{author}{Parent, J.},
  \bibinfo{author}{Civco, D.L.}, \bibinfo{author}{Blei, A.},
  \bibinfo{year}{2017}.
\newblock \bibinfo{title}{The persistent decline in urban densities: Global and
  historical evidence of 'sprawl'}.
\bibitem[{Angel et~al.(2011)Angel, Parent, Civco, Blei and Potere}]{Angel2011}
\bibinfo{author}{Angel, S.}, \bibinfo{author}{Parent, J.},
  \bibinfo{author}{Civco, D.L.}, \bibinfo{author}{Blei, A.},
  \bibinfo{author}{Potere, D.}, \bibinfo{year}{2011}.
\newblock \bibinfo{title}{The dimensions of global urban expansion: Estimates
  and projections for all countries, 2000–2050}.
\newblock \bibinfo{journal}{Progress in Planning} \bibinfo{volume}{75},
  \bibinfo{pages}{53--107}.
\newblock \URLprefix
  \url{https://www.sciencedirect.com/science/article/pii/S0305900611000109},
  \DOIprefix\doi{https://doi.org/10.1016/j.progress.2011.04.001}.
  \bibinfo{note}{the dimensions of global urban expansion: Estimates and
  projections for all countries, 2000–2050}.
\bibitem[{Barrington-Leigh and Millard-Ball(2015)}]{barrington2015century}
\bibinfo{author}{Barrington-Leigh, C.}, \bibinfo{author}{Millard-Ball, A.},
  \bibinfo{year}{2015}.
\newblock \bibinfo{title}{A century of sprawl in the united states}.
\newblock \bibinfo{journal}{Proceedings of the National Academy of Sciences}
  \bibinfo{volume}{112}, \bibinfo{pages}{8244--8249}.
\bibitem[{Barrington-Leigh and Millard-Ball(2020)}]{barrington2020global}
\bibinfo{author}{Barrington-Leigh, C.}, \bibinfo{author}{Millard-Ball, A.},
  \bibinfo{year}{2020}.
\newblock \bibinfo{title}{Global trends toward urban street-network sprawl}.
\newblock \bibinfo{journal}{Proceedings of the National Academy of Sciences}
  \bibinfo{volume}{117}, \bibinfo{pages}{1941--1950}.
\bibitem[{Batty(2006)}]{Batty2006}
\bibinfo{author}{Batty, M.}, \bibinfo{year}{2006}.
\newblock \bibinfo{title}{Rank clocks}.
\newblock \bibinfo{journal}{Nature} \bibinfo{volume}{444},
  \bibinfo{pages}{592--596}.
\newblock \URLprefix \url{https://doi.org/10.1038/nature05302},
  \DOIprefix\doi{10.1038/nature05302}.
\bibitem[{Bettencourt(2013)}]{Bettencourt2013}
\bibinfo{author}{Bettencourt, L.M.A.}, \bibinfo{year}{2013}.
\newblock \bibinfo{title}{The origins of scaling in cities}.
\newblock \bibinfo{journal}{Science} \bibinfo{volume}{340},
  \bibinfo{pages}{1438--1441}.
\newblock \URLprefix
  \url{https://science.sciencemag.org/content/340/6139/1438},
  \DOIprefix\doi{10.1126/science.1235823},
  \href{http://arxiv.org/abs/https://science.sciencemag.org/content/340/6139/1438.full.pdf}{{\tt
  arXiv:https://science.sciencemag.org/content/340/6139/1438.full.pdf}}.
\bibitem[{Bettencourt et~al.(2007)Bettencourt, Lobo, Helbing, K{\"u}hnert and
  West}]{Bettencourt2007}
\bibinfo{author}{Bettencourt, L.M.A.}, \bibinfo{author}{Lobo, J.},
  \bibinfo{author}{Helbing, D.}, \bibinfo{author}{K{\"u}hnert, C.},
  \bibinfo{author}{West, G.B.}, \bibinfo{year}{2007}.
\newblock \bibinfo{title}{Growth, innovation, scaling, and the pace of life in
  cities}.
\newblock \bibinfo{journal}{Proceedings of the National Academy of Sciences}
  \bibinfo{volume}{104}, \bibinfo{pages}{7301--7306}.
\newblock \URLprefix \url{https://www.pnas.org/content/104/17/7301},
  \DOIprefix\doi{10.1073/pnas.0610172104},
  \href{http://arxiv.org/abs/https://www.pnas.org/content/104/17/7301.full.pdf}{{\tt
  arXiv:https://www.pnas.org/content/104/17/7301.full.pdf}}.
\bibitem[{Bettencourt et~al.(2020)Bettencourt, Yang, Lobo, Kempes, Rybski and
  Hamilton}]{Bettencourt2020_longvscross}
\bibinfo{author}{Bettencourt, L.M.A.}, \bibinfo{author}{Yang, V.C.},
  \bibinfo{author}{Lobo, J.}, \bibinfo{author}{Kempes, C.P.},
  \bibinfo{author}{Rybski, D.}, \bibinfo{author}{Hamilton, M.J.},
  \bibinfo{year}{2020}.
\newblock \bibinfo{title}{The interpretation of urban scaling analysis in
  time}.
\newblock \bibinfo{journal}{Journal of The Royal Society Interface}
  \bibinfo{volume}{17}, \bibinfo{pages}{20190846}.
\newblock \URLprefix
  \url{https://royalsocietypublishing.org/doi/abs/10.1098/rsif.2019.0846},
  \DOIprefix\doi{10.1098/rsif.2019.0846},
  \href{http://arxiv.org/abs/https://royalsocietypublishing.org/doi/pdf/10.1098/rsif.2019.0846}{{\tt
  arXiv:https://royalsocietypublishing.org/doi/pdf/10.1098/rsif.2019.0846}}.
\bibitem[{Boeing(2019)}]{Boeing2019efficient}
\bibinfo{author}{Boeing, G.}, \bibinfo{year}{2019}.
\newblock \bibinfo{title}{Urban spatial order: street network orientation,
  configuration, and entropy}.
\newblock \bibinfo{journal}{Applied Network Science} \bibinfo{volume}{4},
  \bibinfo{pages}{67}.
\newblock \URLprefix \url{https://doi.org/10.1007/s41109-019-0189-1},
  \DOIprefix\doi{10.1007/s41109-019-0189-1}.
\bibitem[{Boeing(2020a)}]{Boeing2020multi}
\bibinfo{author}{Boeing, G.}, \bibinfo{year}{2020}a.
\newblock \bibinfo{title}{A multi-scale analysis of 27,000 urban street
  networks: Every us city, town, urbanized area, and zillow neighborhood}.
\newblock \bibinfo{journal}{Environment and Planning B: Urban Analytics and
  City Science} \bibinfo{volume}{47}, \bibinfo{pages}{590--608}.
\newblock \DOIprefix\doi{10.1177/2399808318784595}.
\bibitem[{Boeing(2020b)}]{boeing2020off}
\bibinfo{author}{Boeing, G.}, \bibinfo{year}{2020}b.
\newblock \bibinfo{title}{Off the grid...and back again? the recent evolution
  of american street network planning and design}.
\newblock \bibinfo{journal}{Journal of the American Planning Association}
  \bibinfo{volume}{87}, \bibinfo{pages}{1--15}.
\bibitem[{Bureau(2018)}]{CensusRegions}
\bibinfo{author}{Bureau, C.}, \bibinfo{year}{2018}.
\newblock \bibinfo{title}{cb\_2018\_us\_region\_500k}.
\newblock \URLprefix
  \url{https://www2.census.gov/geo/tiger/GENZ2018/shp/cb\_2018\_us\_region\_500k.zip}.
\bibitem[{Bureau(2020a)}]{census_county_pop_mid}
\bibinfo{author}{Bureau, U.C.}, \bibinfo{year}{2020}a.
\newblock \bibinfo{title}{cenpop2000}.
\bibitem[{Bureau(2020b)}]{census_county_pop_old}
\bibinfo{author}{Bureau, U.C.}, \bibinfo{year}{2020}b.
\newblock \bibinfo{title}{Census u.s. decennial county population data,
  1900-1990}.
\bibitem[{Bureau(2020c)}]{census_county_pop_new}
\bibinfo{author}{Bureau, U.C.}, \bibinfo{year}{2020}c.
\newblock \bibinfo{title}{County population totals: 2010-2019}.
\bibitem[{Bureau(2023)}]{BLM_msa}
\bibinfo{author}{Bureau, U.C.}, \bibinfo{year}{2023}.
\newblock \bibinfo{title}{About}.
\newblock
  \bibinfo{howpublished}{\url{https://www.census.gov/programs-surveys/metro-micro/about.html}}.
\bibitem[{Burghardt and Uhl(2022)}]{Burghardt2022_roadstats}
\bibinfo{author}{Burghardt, K.}, \bibinfo{author}{Uhl, J.H.},
  \bibinfo{year}{2022}.
\newblock \bibinfo{title}{Historical road network statistics for core-based
  statistical areas in the u.s. (1900 - 2010)}.
\newblock \DOIprefix\doi{10.6084/m9.figshare.19584088.v1}.
  \bibinfo{note}{online; accessed 01 January 2023}.
\bibitem[{Burghardt et~al.(2022)Burghardt, Uhl, Lerman and
  Leyk}]{Burghardt2021_city}
\bibinfo{author}{Burghardt, K.}, \bibinfo{author}{Uhl, J.H.},
  \bibinfo{author}{Lerman, K.}, \bibinfo{author}{Leyk, S.},
  \bibinfo{year}{2022}.
\newblock \bibinfo{title}{Road network evolution in the urban and rural united
  states since 1900}.
\newblock \bibinfo{journal}{Computers, Environment and Urban Systems}
  \bibinfo{volume}{95}, \bibinfo{pages}{101803}.
\newblock \URLprefix
  \url{https://www.sciencedirect.com/science/article/pii/S0198971522000473},
  \DOIprefix\doi{10.1016/j.compenvurbsys.2022.101803}.
\bibitem[{Burghardt et~al.(2023)Burghardt, Uhl, Lerman and
  Leyk}]{burghardt2023analyzing}
\bibinfo{author}{Burghardt, K.}, \bibinfo{author}{Uhl, J.H.},
  \bibinfo{author}{Lerman, K.}, \bibinfo{author}{Leyk, S.},
  \bibinfo{year}{2023}.
\newblock \bibinfo{title}{Analyzing urban scaling laws in the united states
  over 115 years}.
\newblock \href{http://arxiv.org/abs/2209.10852}{{\tt arXiv:2209.10852}}.
\bibitem[{Depersin and Barthelemy(2018)}]{Depersin2018}
\bibinfo{author}{Depersin, J.}, \bibinfo{author}{Barthelemy, M.},
  \bibinfo{year}{2018}.
\newblock \bibinfo{title}{From global scaling to the dynamics of individual
  cities}.
\newblock \bibinfo{journal}{Proceedings of the National Academy of Sciences}
  \bibinfo{volume}{115}, \bibinfo{pages}{2317--2322}.
\newblock \URLprefix \url{https://www.pnas.org/content/115/10/2317},
  \DOIprefix\doi{10.1073/pnas.1718690115},
  \href{http://arxiv.org/abs/https://www.pnas.org/content/115/10/2317.full.pdf}{{\tt
  arXiv:https://www.pnas.org/content/115/10/2317.full.pdf}}.
\bibitem[{Deutsch(2001)}]{Deutsch2001encyclopedia}
\bibinfo{author}{Deutsch, C.V.}, \bibinfo{year}{2001}.
\newblock \bibinfo{title}{Geostatistics}.
\newblock \bibinfo{publisher}{Academic Press}.
\bibitem[{Done et~al.(2018)Done, Simmons and Czajkowski}]{Done2018}
\bibinfo{author}{Done, J.M.}, \bibinfo{author}{Simmons, K.M.},
  \bibinfo{author}{Czajkowski, J.}, \bibinfo{year}{2018}.
\newblock \bibinfo{title}{Relationship between residential losses and hurricane
  winds: Role of the florida building code}.
\newblock \bibinfo{journal}{ASCE-ASME Journal of Risk and Uncertainty in
  Engineering Systems, Part A: Civil Engineering} \bibinfo{volume}{4},
  \bibinfo{pages}{04018001}.
\newblock \URLprefix
  \url{https://ascelibrary.org/doi/abs/10.1061/AJRUA6.0000947},
  \DOIprefix\doi{10.1061/AJRUA6.0000947},
  \href{http://arxiv.org/abs/https://ascelibrary.org/doi/pdf/10.1061/AJRUA6.0000947}{{\tt
  arXiv:https://ascelibrary.org/doi/pdf/10.1061/AJRUA6.0000947}}.
\bibitem[{Dunn(1961)}]{Dunn1961}
\bibinfo{author}{Dunn, O.J.}, \bibinfo{year}{1961}.
\newblock \bibinfo{title}{Multiple comparisons among means}.
\newblock \bibinfo{journal}{Journal of the American Statistical Association}
  \bibinfo{volume}{56}, \bibinfo{pages}{52--64}.
\bibitem[{Eeckhout(2004)}]{Eeckhout2004}
\bibinfo{author}{Eeckhout, J.}, \bibinfo{year}{2004}.
\newblock \bibinfo{title}{Gibrat's law for (all) cities}.
\newblock \bibinfo{journal}{American Economic Review} \bibinfo{volume}{94},
  \bibinfo{pages}{1429--1451}.
\newblock \URLprefix
  \url{https://www.aeaweb.org/articles?id=10.1257/0002828043052303},
  \DOIprefix\doi{10.1257/0002828043052303}.
\bibitem[{Fern\'andez-Gracia et~al.(2014)Fern\'andez-Gracia, Suchecki, Ramasco,
  San~Miguel and Egu\'{\i}luz}]{Gracia2014}
\bibinfo{author}{Fern\'andez-Gracia, J.}, \bibinfo{author}{Suchecki, K.},
  \bibinfo{author}{Ramasco, J.J.}, \bibinfo{author}{San~Miguel, M.},
  \bibinfo{author}{Egu\'{\i}luz, V.M.}, \bibinfo{year}{2014}.
\newblock \bibinfo{title}{Is the voter model a model for voters?}
\newblock \bibinfo{journal}{Phys. Rev. Lett.} \bibinfo{volume}{112},
  \bibinfo{pages}{158701}.
\newblock \URLprefix
  \url{https://link.aps.org/doi/10.1103/PhysRevLett.112.158701},
  \DOIprefix\doi{10.1103/PhysRevLett.112.158701}.
\bibitem[{Gabaix and Ioannides(2004)}]{Gabaix2004}
\bibinfo{author}{Gabaix, X.}, \bibinfo{author}{Ioannides, Y.M.},
  \bibinfo{year}{2004}.
\newblock \bibinfo{title}{Chapter 53 - the evolution of city size
  distributions}, in: \bibinfo{editor}{Henderson, J.V.},
  \bibinfo{editor}{Thisse, J.F.} (Eds.), \bibinfo{booktitle}{Cities and
  Geography}. \bibinfo{publisher}{Elsevier}. volume~\bibinfo{volume}{4} of
  \textit{\bibinfo{series}{Handbook of Regional and Urban Economics}}, pp.
  \bibinfo{pages}{2341--2378}.
\newblock \URLprefix
  \url{https://www.sciencedirect.com/science/article/pii/S1574008004800105},
  \DOIprefix\doi{https://doi.org/10.1016/S1574-0080(04)80010-5}.
\bibitem[{Gregory et~al.(2011)Gregory, Johnston, Pratt, Watts and
  Whatmore}]{gregory2011dictionary}
\bibinfo{author}{Gregory, D.}, \bibinfo{author}{Johnston, R.},
  \bibinfo{author}{Pratt, G.}, \bibinfo{author}{Watts, M.},
  \bibinfo{author}{Whatmore, S.}, \bibinfo{year}{2011}.
\newblock \bibinfo{title}{The dictionary of human geography}.
\newblock \bibinfo{publisher}{John Wiley \& Sons}.
\bibitem[{Helmer(2008)}]{BLM_usa}
\bibinfo{author}{Helmer, G.}, \bibinfo{year}{2008}.
\newblock \bibinfo{title}{Micropolitan statistical areas: a few highlights}.
\newblock
  \bibinfo{howpublished}{\url{https://www.bls.gov/opub/mlr/2008/article/micropolitan-statistical-areas-a-few-highlights.htm}}.
\bibitem[{Hu et~al.(2018)Hu, Lin, Xie, Dai and Qui}]{Hu2018}
\bibinfo{author}{Hu, S.}, \bibinfo{author}{Lin, H.}, \bibinfo{author}{Xie, K.},
  \bibinfo{author}{Dai, J.}, \bibinfo{author}{Qui, J.}, \bibinfo{year}{2018}.
\newblock \bibinfo{title}{Impacts of rain and waterlogging on traffic speed and
  volume on urban roads}, in: \bibinfo{booktitle}{2018 21st International
  Conference on Intelligent Transportation Systems (ITSC)}, pp.
  \bibinfo{pages}{2943--2948}.
\newblock \DOIprefix\doi{10.1109/ITSC.2018.8569639}.
\bibitem[{Jiang and Jia(2011)}]{Jiang2011}
\bibinfo{author}{Jiang, B.}, \bibinfo{author}{Jia, T.}, \bibinfo{year}{2011}.
\newblock \bibinfo{title}{Zipf's law for all the natural cities in the united
  states: a geospatial perspective}.
\newblock \bibinfo{journal}{International Journal of Geographical Information
  Science} \bibinfo{volume}{25}, \bibinfo{pages}{1269--1281}.
\newblock \DOIprefix\doi{10.1080/13658816.2010.510801}.
\bibitem[{Keuschnigg(2019)}]{Keuschnigg2019}
\bibinfo{author}{Keuschnigg, M.}, \bibinfo{year}{2019}.
\newblock \bibinfo{title}{Scaling trajectories of cities}.
\newblock \bibinfo{journal}{Proceedings of the National Academy of Sciences}
  \bibinfo{volume}{116}, \bibinfo{pages}{13759--13761}.
\newblock \URLprefix \url{https://www.pnas.org/content/116/28/13759},
  \DOIprefix\doi{10.1073/pnas.1906258116},
  \href{http://arxiv.org/abs/https://www.pnas.org/content/116/28/13759.full.pdf}{{\tt
  arXiv:https://www.pnas.org/content/116/28/13759.full.pdf}}.
\bibitem[{Kruskal and Wallis(1952)}]{Kruskal1952}
\bibinfo{author}{Kruskal, W.H.}, \bibinfo{author}{Wallis, W.A.},
  \bibinfo{year}{1952}.
\newblock \bibinfo{title}{Use of ranks in one-criterion variance analysis}.
\newblock \bibinfo{journal}{Journal of the American Statistical Association}
  \bibinfo{volume}{47}, \bibinfo{pages}{583--621}.
\bibitem[{Lan et~al.(2019)Lan, Li and Zhang}]{Lan2019}
\bibinfo{author}{Lan, T.}, \bibinfo{author}{Li, Z.}, \bibinfo{author}{Zhang,
  H.}, \bibinfo{year}{2019}.
\newblock \bibinfo{title}{Urban allometric scaling beneath structural
  fractality of road networks}.
\newblock \bibinfo{journal}{Annals of the American Association of Geographers}
  \bibinfo{volume}{109}, \bibinfo{pages}{943--957}.
\bibitem[{Lemoy and Caruso(2021)}]{Lemoy2021}
\bibinfo{author}{Lemoy, R.}, \bibinfo{author}{Caruso, G.},
  \bibinfo{year}{2021}.
\newblock \bibinfo{title}{Radial analysis and scaling of urban land use}.
\newblock \bibinfo{journal}{Scientific Reports} \bibinfo{volume}{11},
  \bibinfo{pages}{22044}.
\newblock \URLprefix \url{https://doi.org/10.1038/s41598-021-01477-y},
  \DOIprefix\doi{10.1038/s41598-021-01477-y}.
\bibitem[{Leyk and Uhl(2018a)}]{leyk2018hisdac}
\bibinfo{author}{Leyk, S.}, \bibinfo{author}{Uhl, J.H.}, \bibinfo{year}{2018}a.
\newblock \bibinfo{title}{Hisdac-us, historical settlement data compilation for
  the conterminous united states over 200 years}.
\newblock \bibinfo{journal}{Scientific data} \bibinfo{volume}{5},
  \bibinfo{pages}{180175}.
\bibitem[{Leyk and Uhl(2018b)}]{data_bui2018}
\bibinfo{author}{Leyk, S.}, \bibinfo{author}{Uhl, J.H.}, \bibinfo{year}{2018}b.
\newblock \bibinfo{title}{{Historical built-up intensity layer series for the
  U.S. 1810 - 2015}}.
\newblock \DOIprefix\doi{10.7910/DVN/1WB9E4}.
\bibitem[{Lobo et~al.(2020)Lobo, Bettencourt, Smith and Ortman}]{Lobo2020}
\bibinfo{author}{Lobo, J.}, \bibinfo{author}{Bettencourt, L.M.},
  \bibinfo{author}{Smith, M.E.}, \bibinfo{author}{Ortman, S.},
  \bibinfo{year}{2020}.
\newblock \bibinfo{title}{Settlement scaling theory: Bridging the study of
  ancient and contemporary urban systems}.
\newblock \bibinfo{journal}{Urban Studies} \bibinfo{volume}{57},
  \bibinfo{pages}{731--747}.
\newblock \DOIprefix\doi{10.1177/0042098019873796}.
\bibitem[{Louf and Barthelemy(2014)}]{Louf2014}
\bibinfo{author}{Louf, R.}, \bibinfo{author}{Barthelemy, M.},
  \bibinfo{year}{2014}.
\newblock \bibinfo{title}{How congestion shapes cities: from mobility patterns
  to scaling}.
\newblock \bibinfo{journal}{Scientific Reports} \bibinfo{volume}{4},
  \bibinfo{pages}{5561}.
\newblock \URLprefix \url{https://doi.org/10.1038/srep05561},
  \DOIprefix\doi{10.1038/srep05561}.
\bibitem[{Microsoft(2020)}]{MBF2020}
\bibinfo{author}{Microsoft}, \bibinfo{year}{2020}.
\newblock \bibinfo{title}{Usbuildingfootprints}.
\newblock \URLprefix \url{https://github.com/Microsoft/USBuildingFootprints}.
\bibitem[{Ribeiro and Rybski(2023)}]{Ribeiro2023}
\bibinfo{author}{Ribeiro, F.L.}, \bibinfo{author}{Rybski, D.},
  \bibinfo{year}{2023}.
\newblock \bibinfo{title}{Mathematical models to explain the origin of urban
  scaling laws}.
\newblock \bibinfo{journal}{Physics Reports} \bibinfo{volume}{1012},
  \bibinfo{pages}{1--39}.
\newblock \URLprefix
  \url{https://www.sciencedirect.com/science/article/pii/S0370157323000650},
  \DOIprefix\doi{https://doi.org/10.1016/j.physrep.2023.02.002}.
\bibitem[{Rozenfeld et~al.(2008)Rozenfeld, Rybski, Andrade, Batty, Stanley and
  Makse}]{Rozenfeld2008}
\bibinfo{author}{Rozenfeld, H.D.}, \bibinfo{author}{Rybski, D.},
  \bibinfo{author}{Andrade, J.S.}, \bibinfo{author}{Batty, M.},
  \bibinfo{author}{Stanley, H.E.}, \bibinfo{author}{Makse, H.A.},
  \bibinfo{year}{2008}.
\newblock \bibinfo{title}{Laws of population growth}.
\newblock \bibinfo{journal}{Proceedings of the National Academy of Sciences}
  \bibinfo{volume}{105}, \bibinfo{pages}{18702--18707}.
\newblock \URLprefix \url{https://www.pnas.org/content/105/48/18702},
  \DOIprefix\doi{10.1073/pnas.0807435105},
  \href{http://arxiv.org/abs/https://www.pnas.org/content/105/48/18702.full.pdf}{{\tt
  arXiv:https://www.pnas.org/content/105/48/18702.full.pdf}}.
\bibitem[{Rozenfeld et~al.(2011)Rozenfeld, Rybski, Gabaix and
  Makse}]{Rozenfeld2011}
\bibinfo{author}{Rozenfeld, H.D.}, \bibinfo{author}{Rybski, D.},
  \bibinfo{author}{Gabaix, X.}, \bibinfo{author}{Makse, H.A.},
  \bibinfo{year}{2011}.
\newblock \bibinfo{title}{The area and population of cities: New insights from
  a different perspective on cities}.
\newblock \bibinfo{journal}{American Economic Review} \bibinfo{volume}{101},
  \bibinfo{pages}{2205–2225}.
\bibitem[{Samaniego and Moses(2008)}]{Samaniego2008}
\bibinfo{author}{Samaniego, H.}, \bibinfo{author}{Moses, M.E.},
  \bibinfo{year}{2008}.
\newblock \bibinfo{title}{Cities as organisms: Allometric scaling of urban road
  networks}.
\newblock \bibinfo{journal}{Journal of Transport and Land Use}
  \bibinfo{volume}{1}.
\newblock \URLprefix \url{https://www.jtlu.org/index.php/jtlu/article/view/29},
  \DOIprefix\doi{10.5198/jtlu.v1i1.29}.
\bibitem[{Strano et~al.(2017)Strano, Giometto, Shai, Bertuzzo, Mucha and
  Rinaldo}]{Strano2017}
\bibinfo{author}{Strano, E.}, \bibinfo{author}{Giometto, A.},
  \bibinfo{author}{Shai, S.}, \bibinfo{author}{Bertuzzo, E.},
  \bibinfo{author}{Mucha, P.J.}, \bibinfo{author}{Rinaldo, A.},
  \bibinfo{year}{2017}.
\newblock \bibinfo{title}{The scaling structure of the global road network}.
\newblock \bibinfo{journal}{Royal Society Open Science} \bibinfo{volume}{4},
  \bibinfo{pages}{170590}.
\newblock \DOIprefix\doi{10.1098/rsos.170590}.
\bibitem[{Uhl and Burghardt(2022)}]{Uhl2022_urbanarea}
\bibinfo{author}{Uhl, J.H.}, \bibinfo{author}{Burghardt, K.},
  \bibinfo{year}{2022}.
\newblock \bibinfo{title}{Historical, generalized built-up areas in u.s.
  core-based statistical areas 1900 - 2015}.
\newblock
  \bibinfo{howpublished}{\url{https://doi.org/10.6084/m9.figshare.19593409.v2}}.
\bibitem[{Uhl and Leyk(2020a)}]{databua2020}
\bibinfo{author}{Uhl, J.H.}, \bibinfo{author}{Leyk, S.}, \bibinfo{year}{2020}a.
\newblock \bibinfo{title}{{Historical built-up areas (BUA) - gridded surfaces
  for the U.S. from 1810 to 2015}}.
\newblock \DOIprefix\doi{10.7910/DVN/J6CYUJ}.
\bibitem[{Uhl and Leyk(2020b)}]{databupl2020}
\bibinfo{author}{Uhl, J.H.}, \bibinfo{author}{Leyk, S.}, \bibinfo{year}{2020}b.
\newblock \bibinfo{title}{{Historical built-up property locations (BUPL) -
  gridded surfaces for the U.S. from 1810 to 2015}}.
\newblock \DOIprefix\doi{10.7910/DVN/SJ213V}.
\bibitem[{Uhl and Leyk(2020c)}]{databupr2020}
\bibinfo{author}{Uhl, J.H.}, \bibinfo{author}{Leyk, S.}, \bibinfo{year}{2020}c.
\newblock \bibinfo{title}{Historical built-up property records (bupr) - gridded
  surfaces for the u.s. from 1810 to 2015}.
\newblock \DOIprefix\doi{10.7910/DVN/YSWMDR}.
\bibitem[{Uhl and Leyk(2022)}]{BUFA}
\bibinfo{author}{Uhl, J.H.}, \bibinfo{author}{Leyk, S.}, \bibinfo{year}{2022}.
\newblock \bibinfo{title}{{Historical building footprint area (BUFA) - gridded
  surfaces for the conterminous U.S. from 1900 to 2010}}.
\newblock \DOIprefix\doi{10.7910/DVN/HXQWNJ}.
\bibitem[{U.S. Geological~Survey(2018)}]{NTD2020}
\bibinfo{author}{U.S. Geological~Survey, N.G.T.O.C.}, \bibinfo{year}{2018}.
\newblock \bibinfo{title}{Usgs national transportation dataset (ntd)}.
\newblock \URLprefix
  \url{https://thor-f5.er.usgs.gov/ngtoc/metadata/waf/transportation/ntd/}.
\bibitem[{Zipf(1949)}]{Zipf1949}
\bibinfo{author}{Zipf, G.K.}, \bibinfo{year}{1949}.
\newblock \bibinfo{title}{Human behavior and the principle of least effort}.
\newblock \bibinfo{publisher}{Addison-Wesley Press}.

\end{thebibliography}

\setcounter{table}{0}
\setcounter{figure}{0}
\renewcommand\thefigure{S\arabic{figure}}    
\renewcommand\thetable{S\arabic{table}}    

\newpage
\section*{Supplementary Information}
\subsection*{Generalized city growth scaling}

\begin{table*}[hb]
    \centering
    \footnotesize
    \begin{tabular}{|p{4.0cm}|p{0.9cm}|p{7.2cm}|}
 \hline
\textbf{Road network metric} &
\textbf{Unit} & \textbf{Description}
\\\hline\hline
Local griddedness & Node &  Number of quadrilaterals touching a node divided by its degree (\citenum{Burghardt2021_city})\\\hline
Road density & Edge	 &  km road per km built-up area\\\hline
Orientation entropy & Edge & Entropy of edge orientation angles, discretized into bins of $5^{\circ}$ (\citenum{Burghardt2021_city,Boeing2019efficient})\\\hline
Mean degree & Node &  Mean number of edges per intersection\\\hline
Dead end rate& Node & Percentage of nodes of degree 1 (\citenum{Burghardt2021_city})\\\hline
Percentage degree 4+& Node & Percentage of nodes with degree 4 or higher (\citenum{Burghardt2021_city})\\\hline
 \hline
    \end{tabular}
    \caption{Urban statistics used to compare city growth correlations.}
    \label{tab:growth_descriptions}
\end{table*}

\begin{figure}[tbh!]
    \centering
    \includegraphics[width=\linewidth]{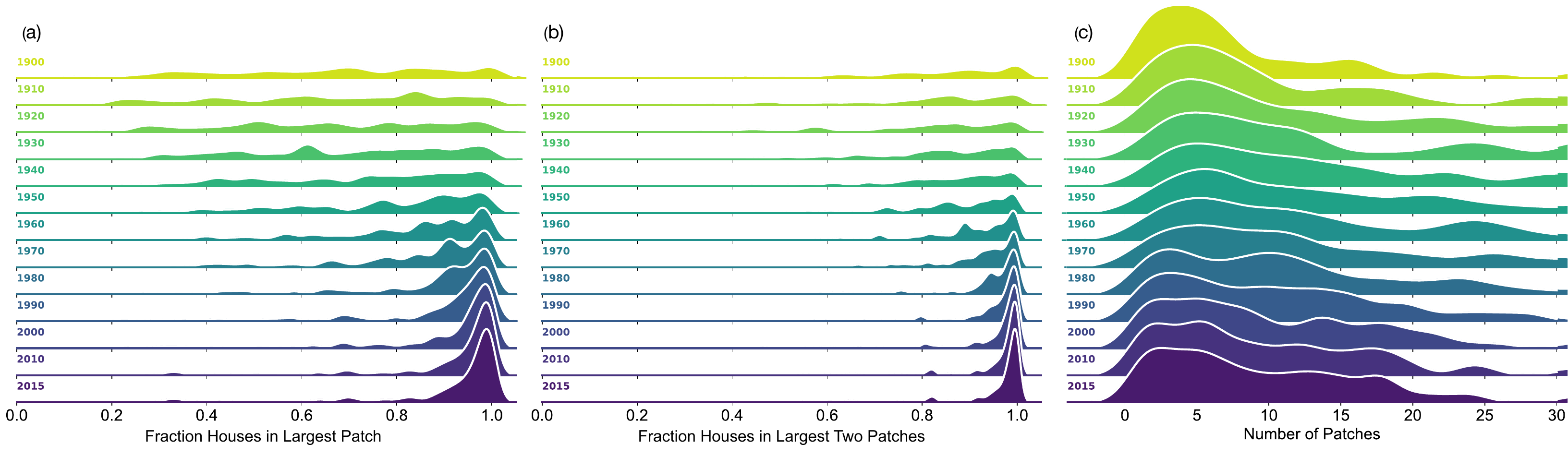}
    \caption{Size of the largest patches. (a) Fraction of houses in the single largest patch within a CBSA. (b) Fraction of houses in the top two largest patches within a CBSA. (c) Patch distribution over time.}
    \label{fig:largest_patches}
\end{figure}

\begin{figure*}
    \centering
    \includegraphics[width=0.6\linewidth]{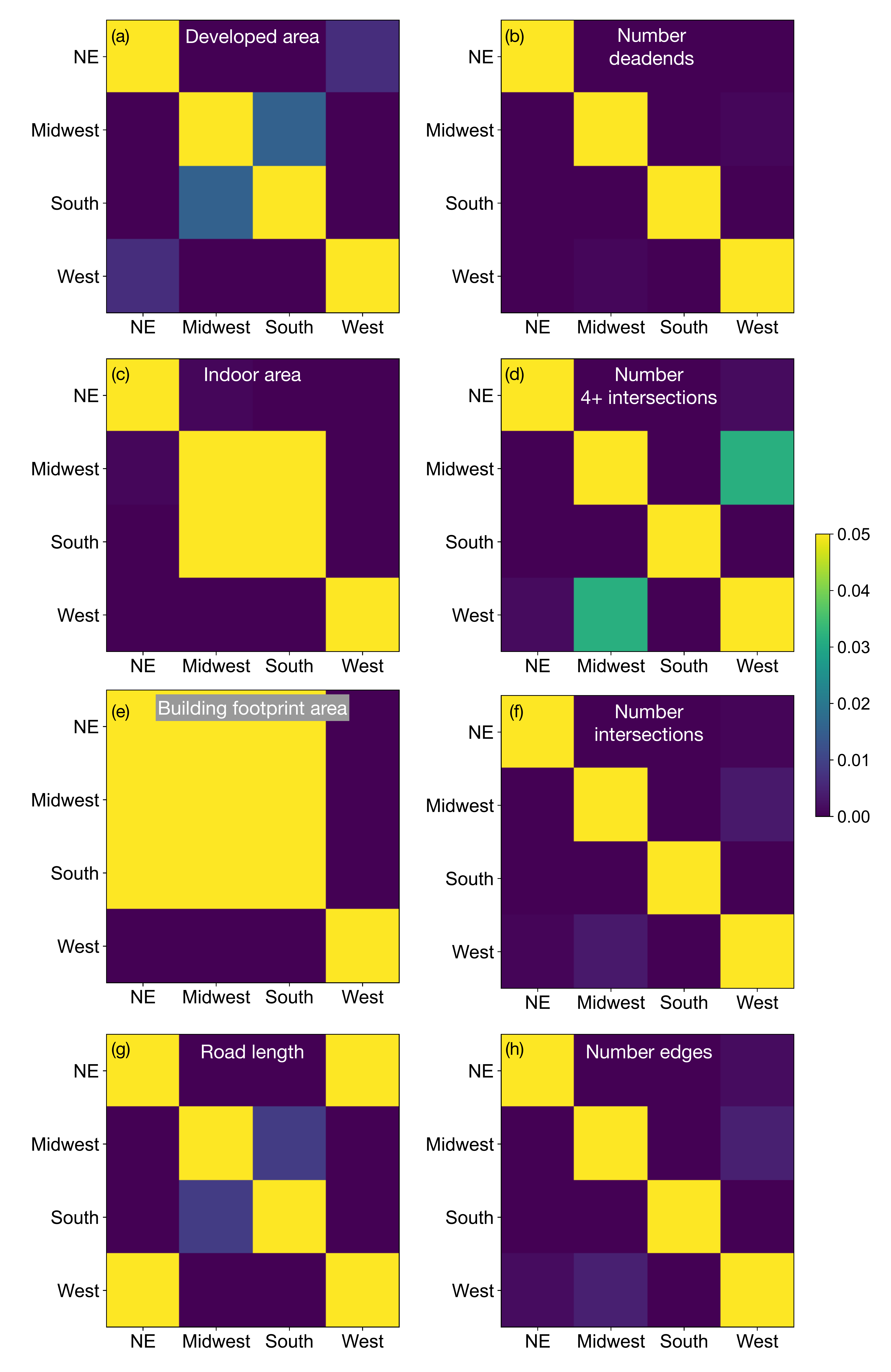}
    \caption{Statistical significance of temporal scaling law distributions within each city size and region for temporal completeness greater than 60\% and spatial coverage greater than 40\% (see main text Fig. 4 for the distributions). Colors correspond to insignificant differences (dark) and statistically significant differences (light) in these statistics, based on Dunn’s test (\cite{Dunn1961}) after rejection by the Kruskal-Wallis test (p-value $< 0.05$) (\cite{Kruskal1952}). Statistically significant differences for (a) developed area, (b) number of deadends, (c) indoor area, (d) number of intersections where four or more edges meet, (e) building footprint area, (f) total number of intersections, (g) road length, and (h) number of edges. }
    \label{fig:scaling_sig_region}
\end{figure*}

\begin{figure*}
    \centering
    \includegraphics[width=0.6\linewidth]{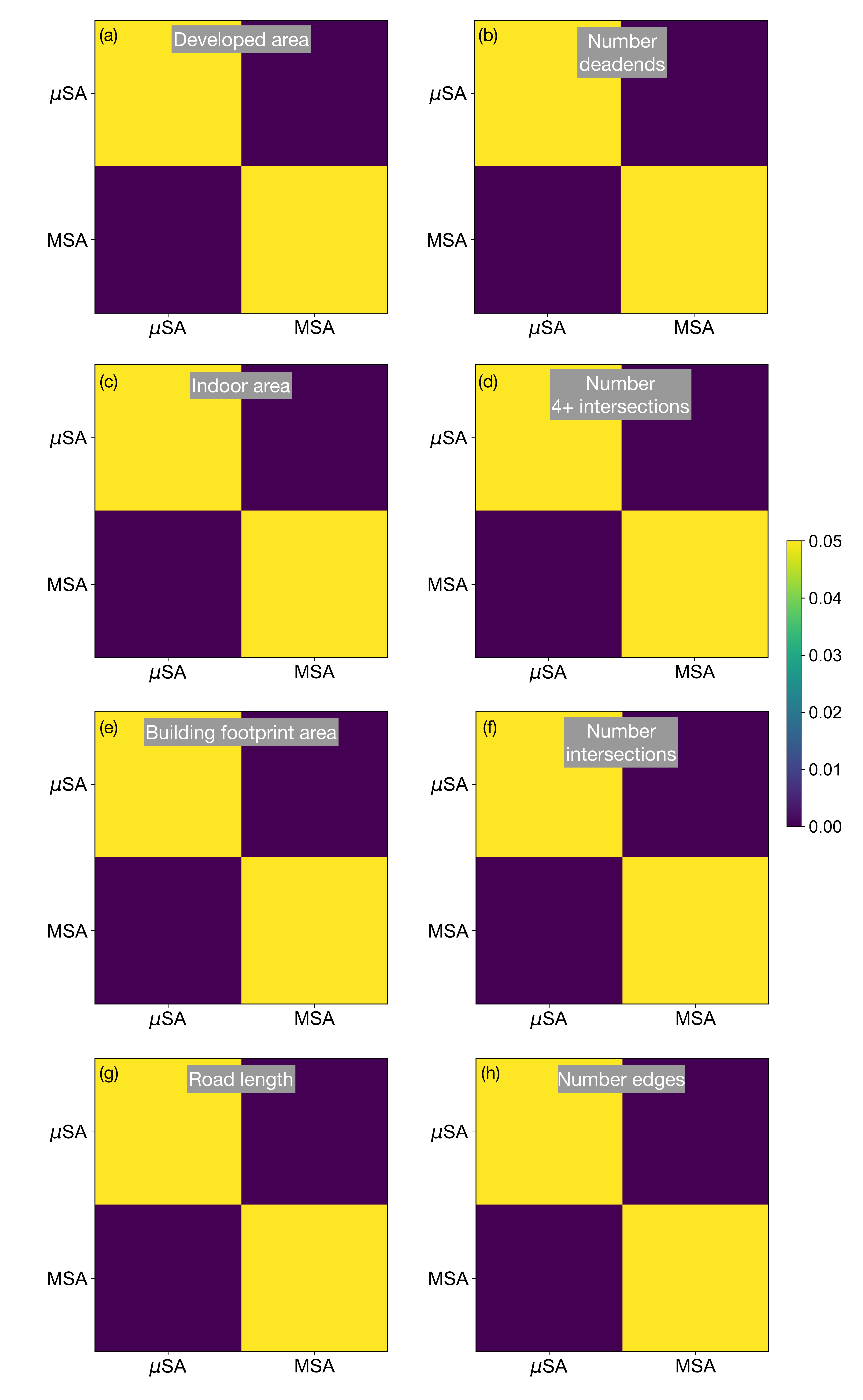}
    \caption{
    Statistical significance of temporal scaling law distributions within each city size and region for temporal completeness greater than 60\% and spatial coverage greater than 40\% (see main text Fig. 4 for the distributions). Colors correspond to insignificant differences (dark) and statistically significant differences (light) in these statistics, based on Dunn’s test (\cite{Dunn1961}) after rejection by the Kruskal-Wallis test (p-value $< 0.05$) (\cite{Kruskal1952}). Statistically significant differences between micro- and metropolitan statistical areas scaling law exponents for (a) developed area, (b) number of deadends, (c) indoor area, (d) number of intersections where four or more edges meet, (e) building footprint area, (f) total number of intersections, (g) road length, and (h) number of edges.  
    }
    \label{fig:scaling_sig_sa}
\end{figure*}


City growth need not be strictly looked at from the lens of city scaling, however. Not only are there a number of ways to define scaling laws (\citenum{Lemoy2021}), we cannot be certain that cities strictly follow temporal scaling laws. We therefore want to check the robustness of these correlations without scaling assumptions. To this end, we analyze the correlations between the growth patterns of any pair of cities. We specifically plot the change in various statistics over time, such as new developed area each year and calculate the Spearman correlation between all cities within a small distance window. As before, we vary the window to plot how correlations decrease with distance between cities. The statistics we calculate are developed area, indoor area, building footprint area, road length (just as when calculating scaling laws), as well as road network statistics outlined in Table~\ref{tab:descriptions}. We calculate the local griddedness (defined in ~(\citenum{Burghardt2021_city}), road density (length of road per square kilometer), orientation entropy (variability in road orientations, defined in (\citenum{boeing2020off}), mean degree (number of edges at an intersection), dead end percentage (the proportion of cul-de-sacs and other roads that do not end at an intersection), and percentage degree 4+ (the proportion of intersections where four or more edges meet). Descriptions of statistics not mentioned in the main text are shown in SI Table~\ref{tab:growth_descriptions}.

\begin{figure*}
    \includegraphics[width=0.9\linewidth]{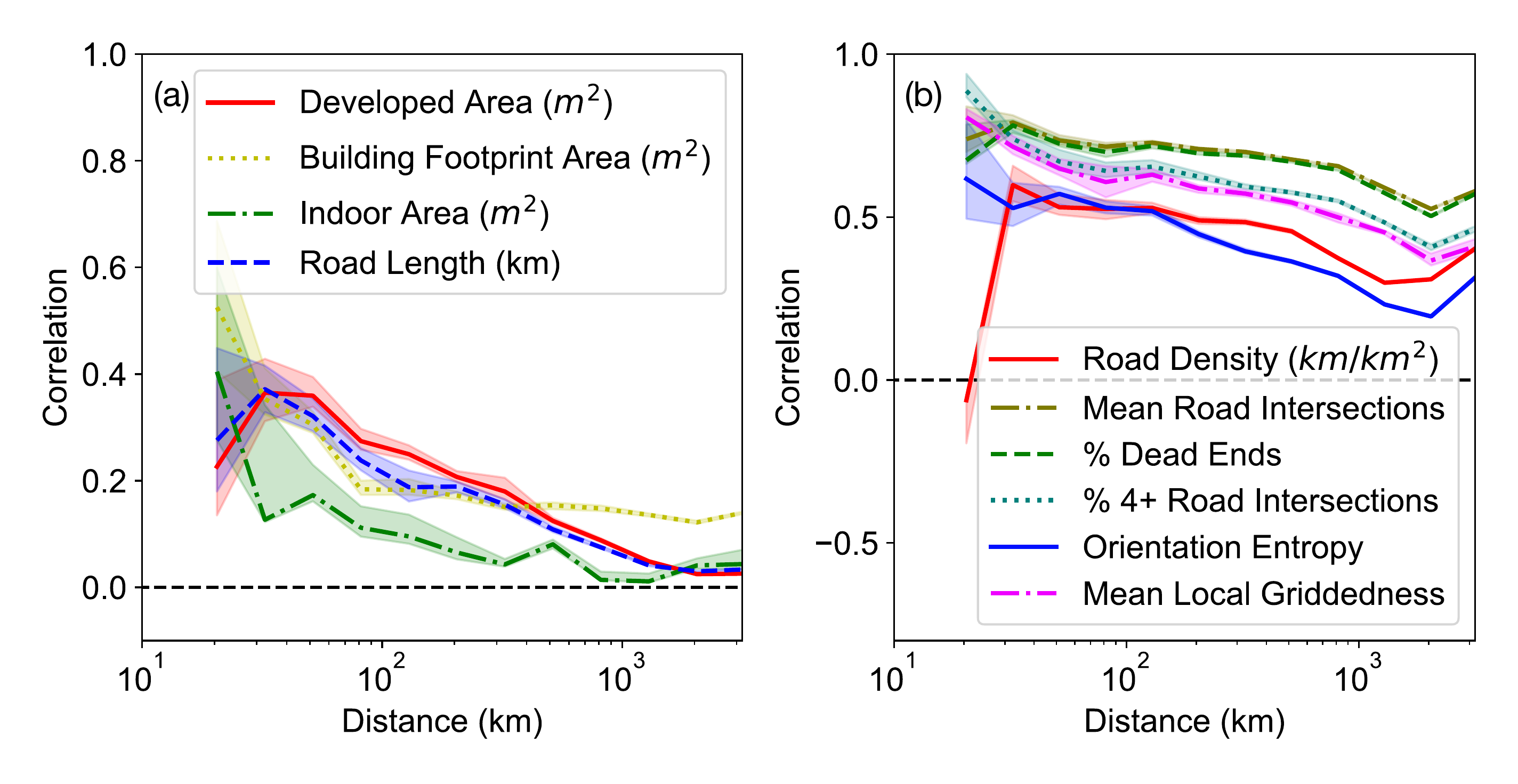}
    \caption{Nearby cities grow similarly. (a) Correlations of extrinsic city properties: developed area growth, building footprint area growth, indoor area growth, and road length growth evolution versus city distance as of 2015.
    road networks growth between cities (CBSAs). (b) Correlations of road network properties: road density, mean number of intersections, fraction of dead ends, orientation entropy (\citenum{Boeing2019efficient}), and mean local griddedness (\citenum{Burghardt2021_city}) of new roads.}
    \label{fig:growth_correl_all}
\end{figure*}

\begin{figure*}
    \includegraphics[width=0.9\linewidth]{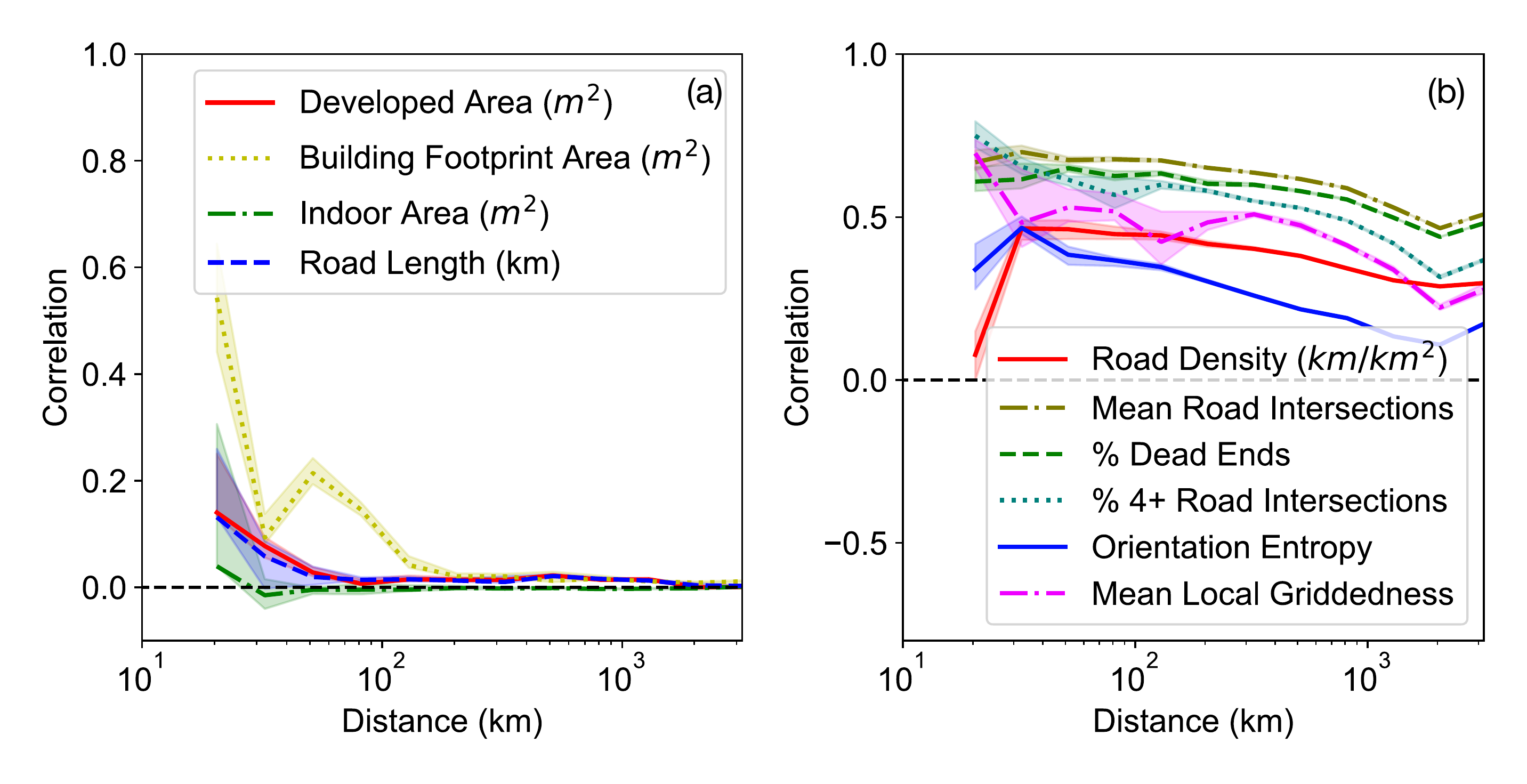}
    \caption{Nearby cities grow similarly. Figures are the same as main text Fig. 6 except we include all CBSAs with temporal completeness and geospatial coverage greater than 0\%. (a) Correlations of extrinsic city properties: developed area growth, building footprint area growth, indoor area growth, and road length growth evolution versus city distance as of 2015.
    road networks growth between cities (CBSAs). (b) Correlations of road network properties: road density, mean number of intersections, fraction of dead ends, orientation entropy (\citenum{Boeing2019efficient}), and mean local griddedness (\citenum{Burghardt2021_city} of new roads.}
    \label{fig:growth_correl_00}
\end{figure*}

\begin{figure*}
    \includegraphics[width=0.9\linewidth]{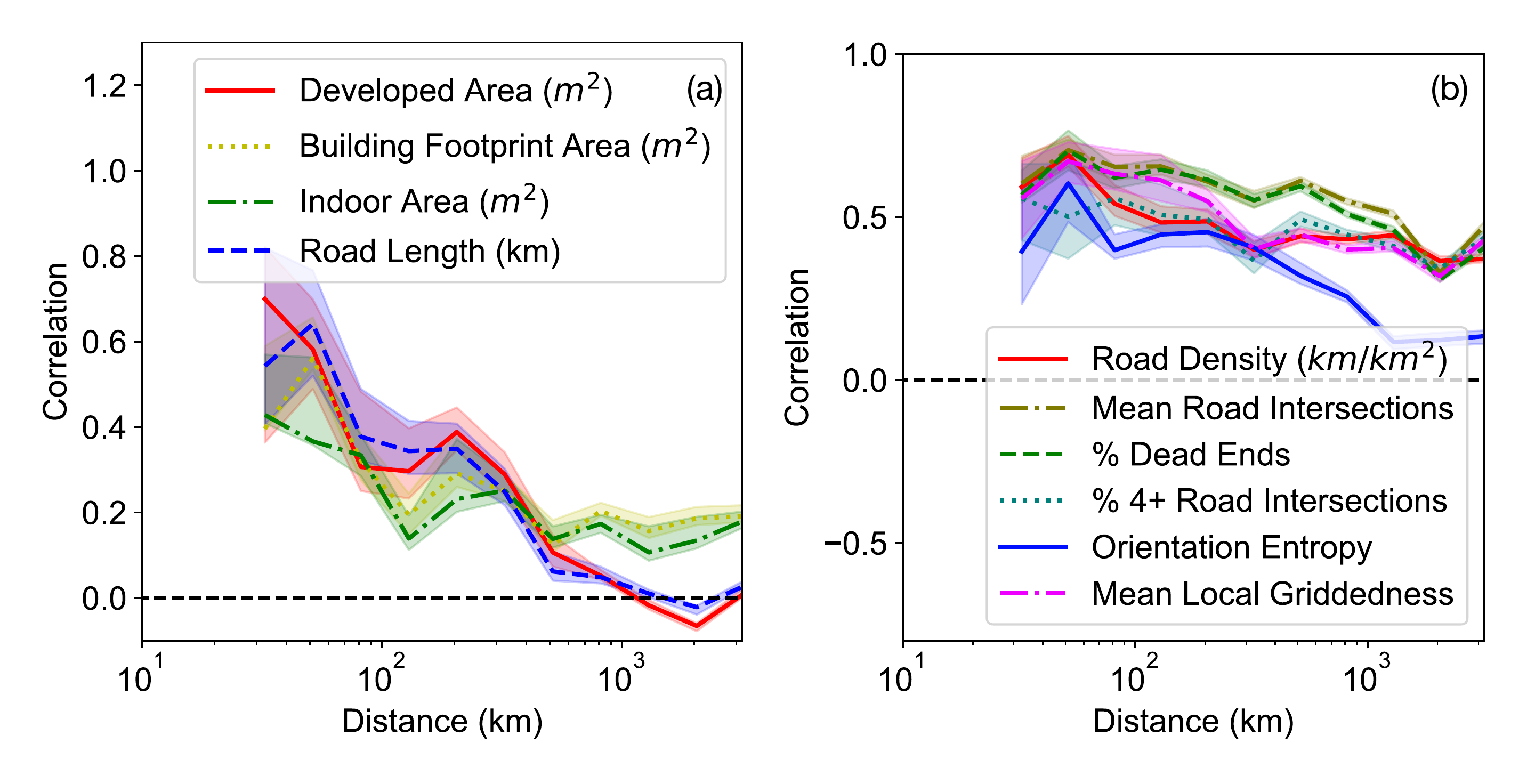}
    \caption{Nearby cities grow similarly. Figures are the same as main text Fig. 6 except we include all CBSAs with temporal completeness and geospatial coverage greater than 80\%. (a) Correlations of extrinsic city properties: developed area growth, building footprint area growth, indoor area growth, and road length growth evolution versus city distance as of 2015.
    road networks growth between cities (CBSAs). (b) Correlations of road network properties: road density, mean number of intersections, fraction of dead ends, orientation entropy (\citenum{Boeing2019efficient}), and mean local griddedness (\citenum{Burghardt2021_city} of new roads.}
    \label{fig:growth_correl_8080}
\end{figure*}

\begin{figure*}
    \centering
    \includegraphics[width=\linewidth]{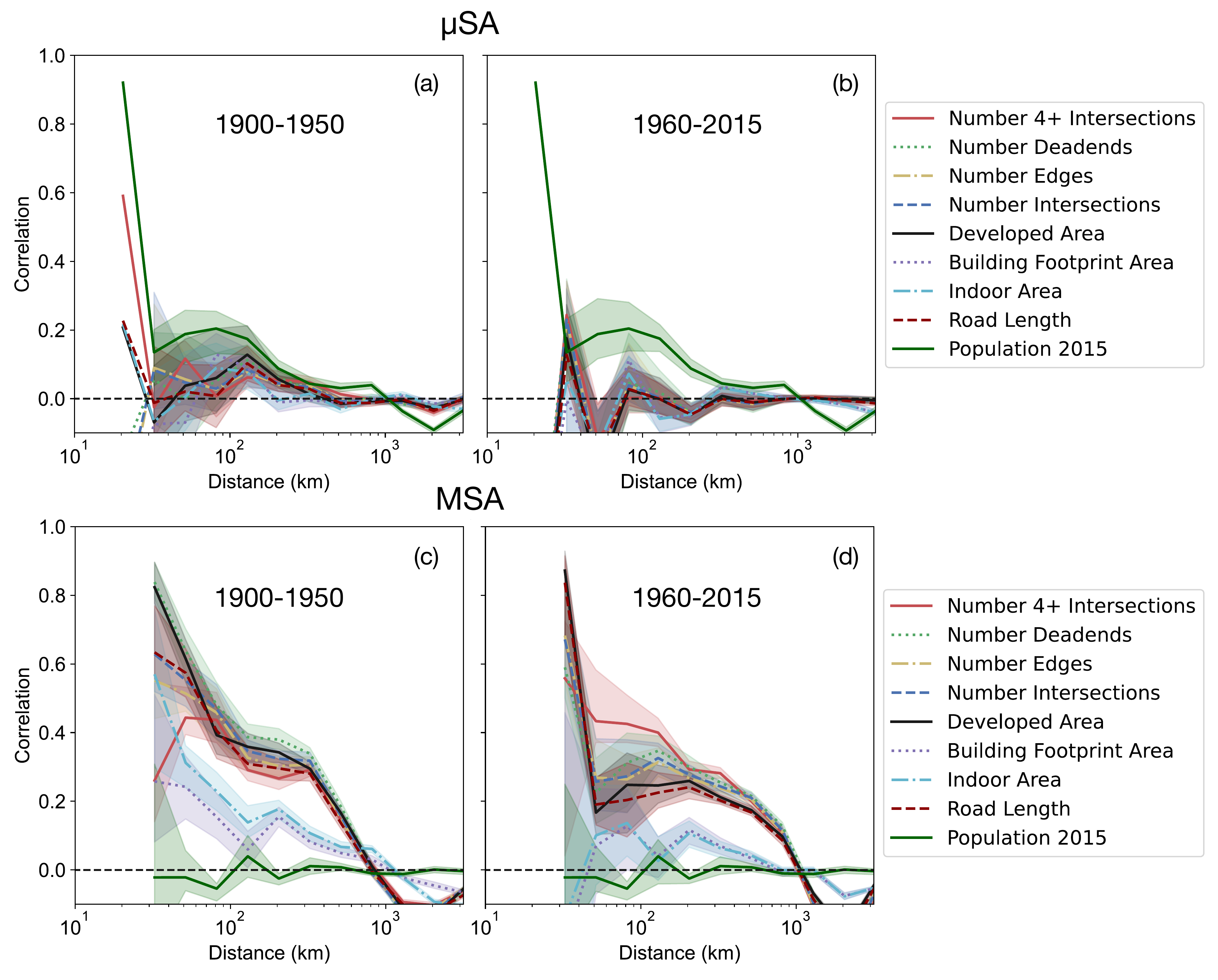}
   \caption{Nearby cities scale similarly. Figures are the same as main text Fig. 6 except we split data by (a--b) MSA and (c--d) $\mu$SA with borders defined as of 2010 and we divide the temporal scaling laws to be between (a,c) 1900--1950 and (b,d) 1960--2015. Spearman correlation versus distance for the number of 4+ intersections, number of deadends, number of edges, number of intersections, developed area, building footprint area, indoor area, and road length. Shaded regions represent 68\% confidence intervals in the mean correlation. }
    \label{fig:growth_correl_msa-usa}
\end{figure*}

The results are shown in Fig.~\ref{fig:growth_correl_all}. In Fig.~\ref{fig:growth_correl_all}a, we see changes in extrinsic variables, namely developed area, building footprint area, indoor area, and road length, all have similarly long-range correlations with city distance. They only reach approximately zero correlation after 1000km, although the decrease in correlation appears faster than scaling law exponents. Figure~\ref{fig:growth_correl_all}b, meanwhile, shows that changes in intrinsic variables (road network statistics) show a qualitative decrease in correlation with distance, although the correlations are always high and never reach zero. This is probably because all cities, regardless of population, experienced similar road network changes with the advent of cars, as described in prior work (\citenum{Burghardt2021_city}). All results are found to be robust when splitting by MSAs and $\mu$SAs (SI Fig.~\ref{fig:growth_correl_msa-usa}), as well as weaker and stronger data filtering (SI Figs.~\ref{fig:growth_correl_00} \&~\ref{fig:growth_correl_8080}, respectively).
These results broadly match the findings seen for scaling laws, suggesting the growth and evolution of cities has a long-range correlation.

\begin{figure*}
    \centering
    \includegraphics[width=\linewidth]{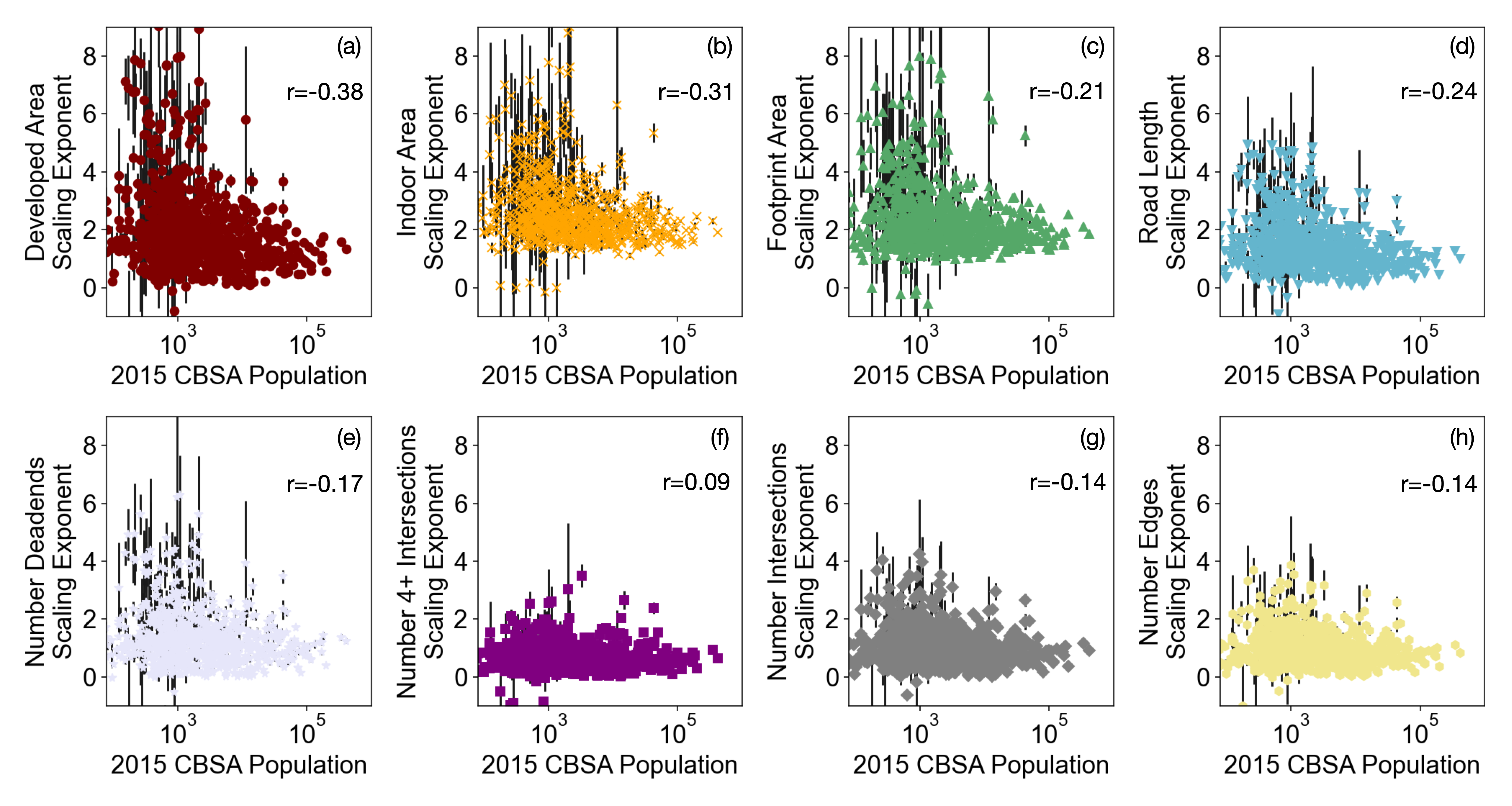}
    \caption{Scaling law exponent versus 2015 population for all CBSAs with temporal completeness and geospatial coverage $>0\%$. (a) Developed area, (b) indoor area, (c) building footprint area, (d) road length, (e) number of deadends, (f) number of intersections with four or more edges meeting, (g) total number of intersections, and (h) number of edges. Black error bars represent standard errors of scaling law coefficients. Statistically significant Spearman correlations (p-values$<0.05$) are shown in each figure, and are otherwise denoted ``n.s.'' (not significant).}
    \label{fig:scalingvspop_00}
\end{figure*}

\begin{figure*}
    \centering
    \includegraphics[width=\linewidth]{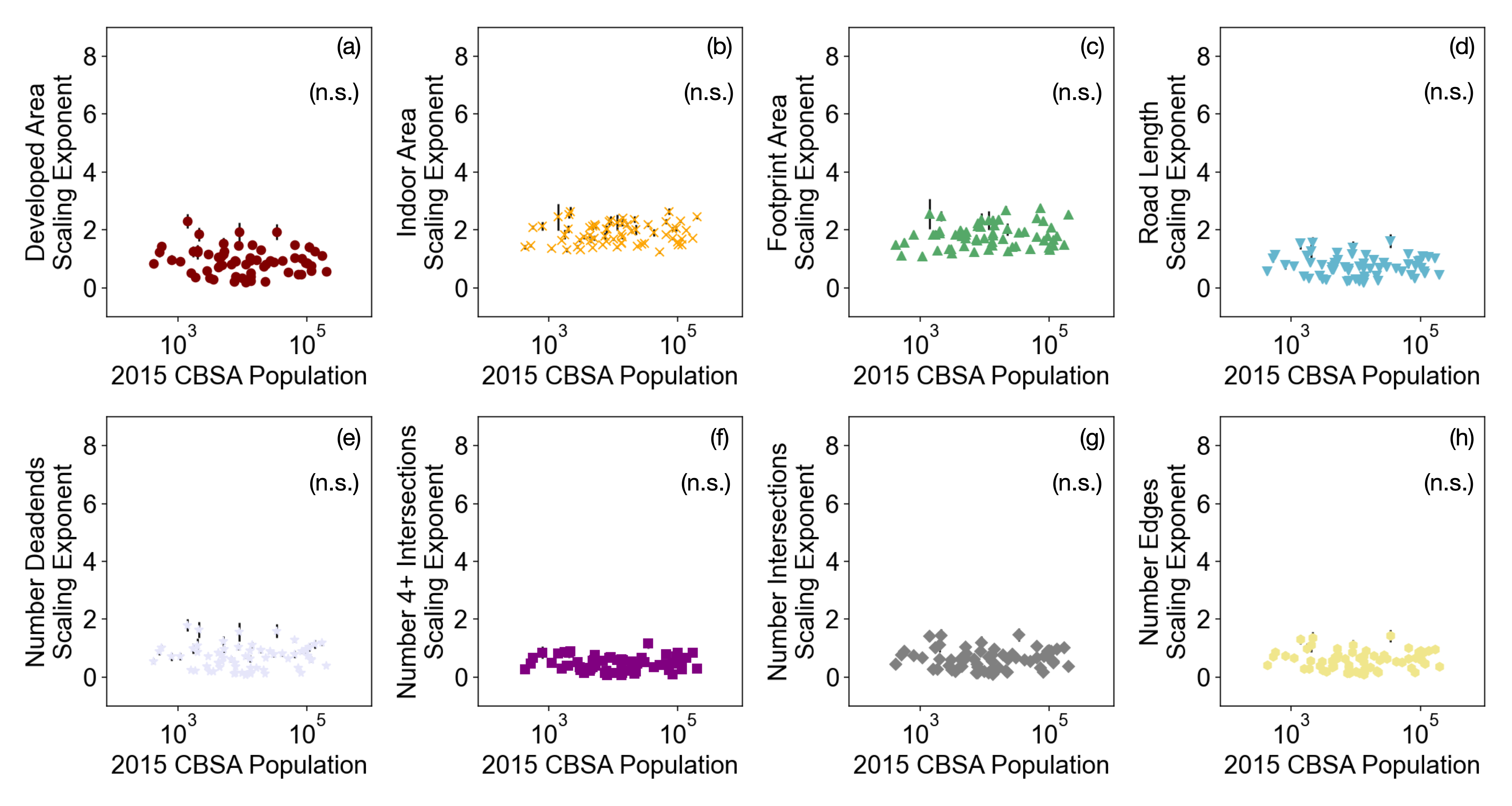}
    \caption{Scaling law exponent versus 2015 population for all CBSAs with temporal completeness and geospatial coverage $>80\%$. (a) Developed area, (b) indoor area, (c) building footprint area, (d) road length, (e) number of deadends, (f) number of intersections with four or more edges meeting, (g) total number of intersections, and (h) number of edges. Black error bars represent standard errors of scaling law coefficients. Statistically significant Spearman correlations (p-values$<0.05$) are shown in each figure, and are otherwise denoted ``n.s.'' (not significant).}
    \label{fig:scalingvspop_8080}
\end{figure*}

\begin{figure*}
    \centering
    \includegraphics[width=\linewidth]{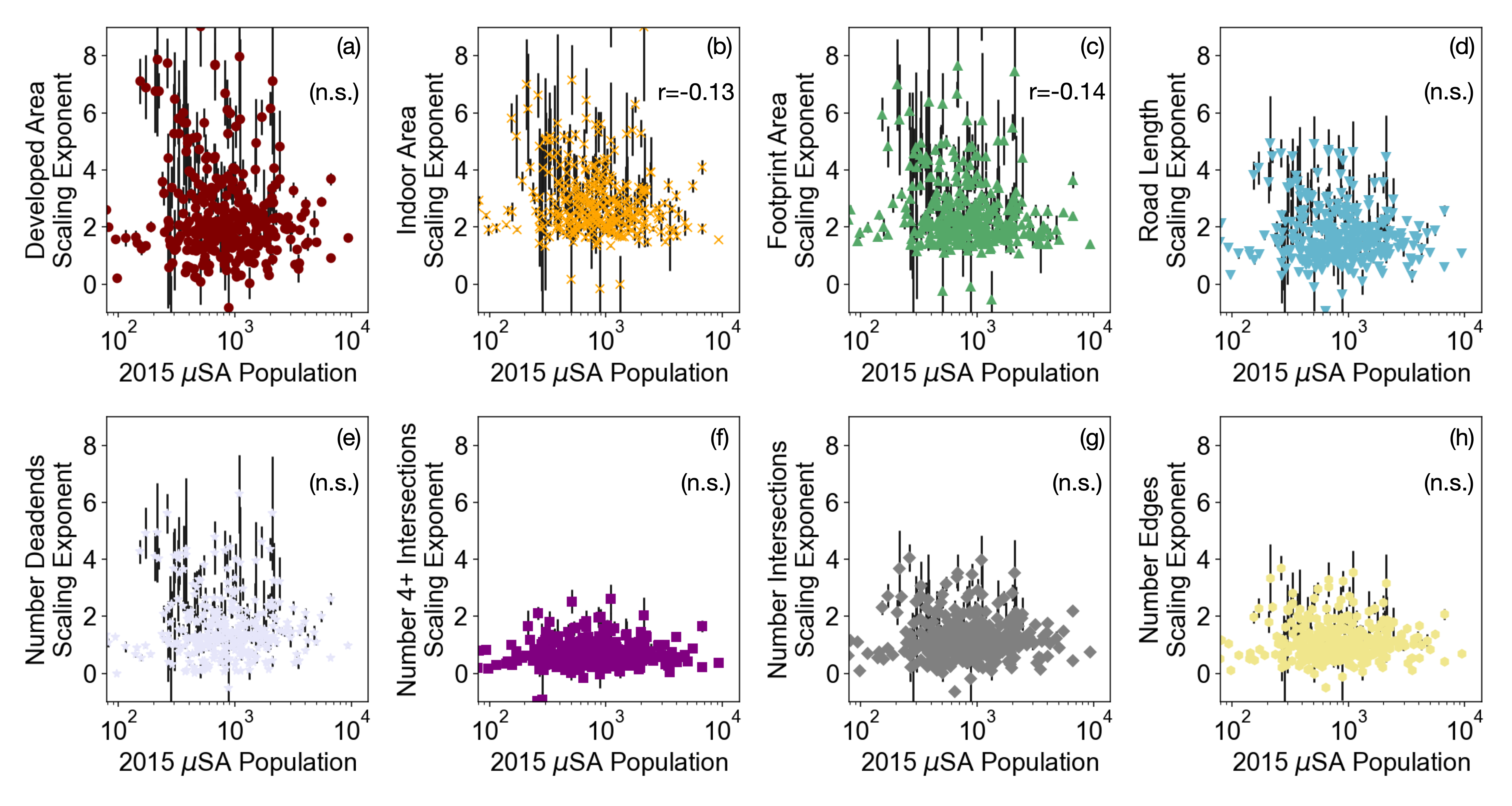}
    \caption{Scaling law exponent versus 2015 population for $\mu$SAs with temporal completeness $>60\%$ and geospatial coverage $>40\%$ as in the main text. (a) Developed area, (b) indoor area, (c) building footprint area, (d) road length, (e) number of deadends, (f) number of intersections with four or more edges meeting, (g) total number of intersections, and (h) number of edges. Black error bars represent standard errors of scaling law coefficients. Statistically significant Spearman correlations (p-values$<0.05$) are shown in each figure, and are otherwise denoted ``n.s.'' (not significant).}
    \label{fig:scalingvspop_usa}
\end{figure*}

\begin{figure*}
    \centering
    \includegraphics[width=\linewidth]{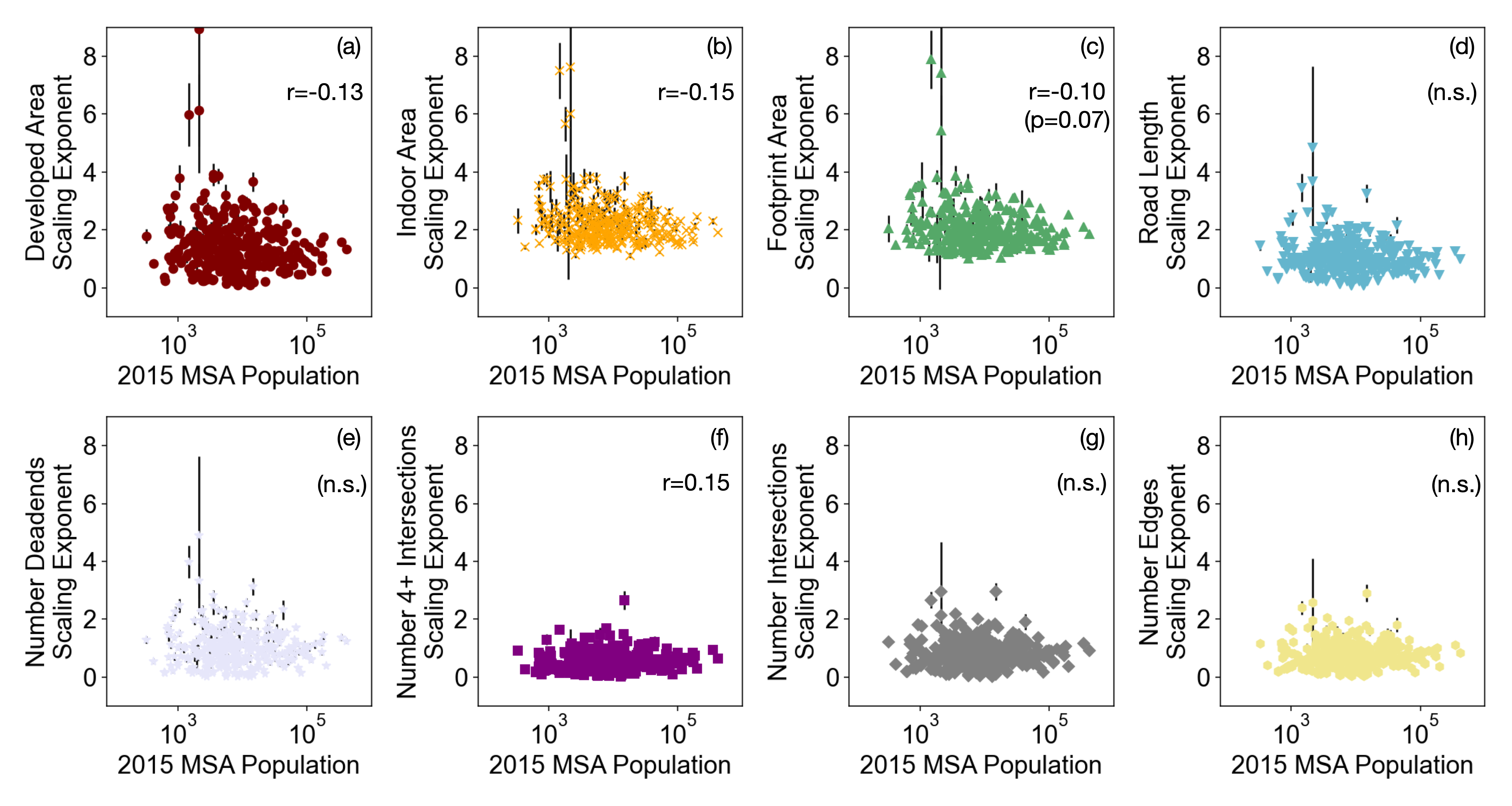}
    \caption{Scaling law exponent versus 2015 population for MSAs with temporal completeness $>60\%$ and geospatial coverage $>40\%$ as in the main text. (a) Developed area, (b) indoor area, (c) building footprint area, (d) road length, (e) number of deadends, (f) number of intersections with four or more edges meeting, (g) total number of intersections, and (h) number of edges. Black error bars represent standard errors of scaling law coefficients. Statistically significant Spearman correlations (p-values$<0.05$) are shown in each figure, and are otherwise denoted ``n.s.'' (not significant).}
    \label{fig:scalingvspop_MSA}
\end{figure*}

\begin{figure*}
    \centering
    \includegraphics[width=0.7\linewidth]{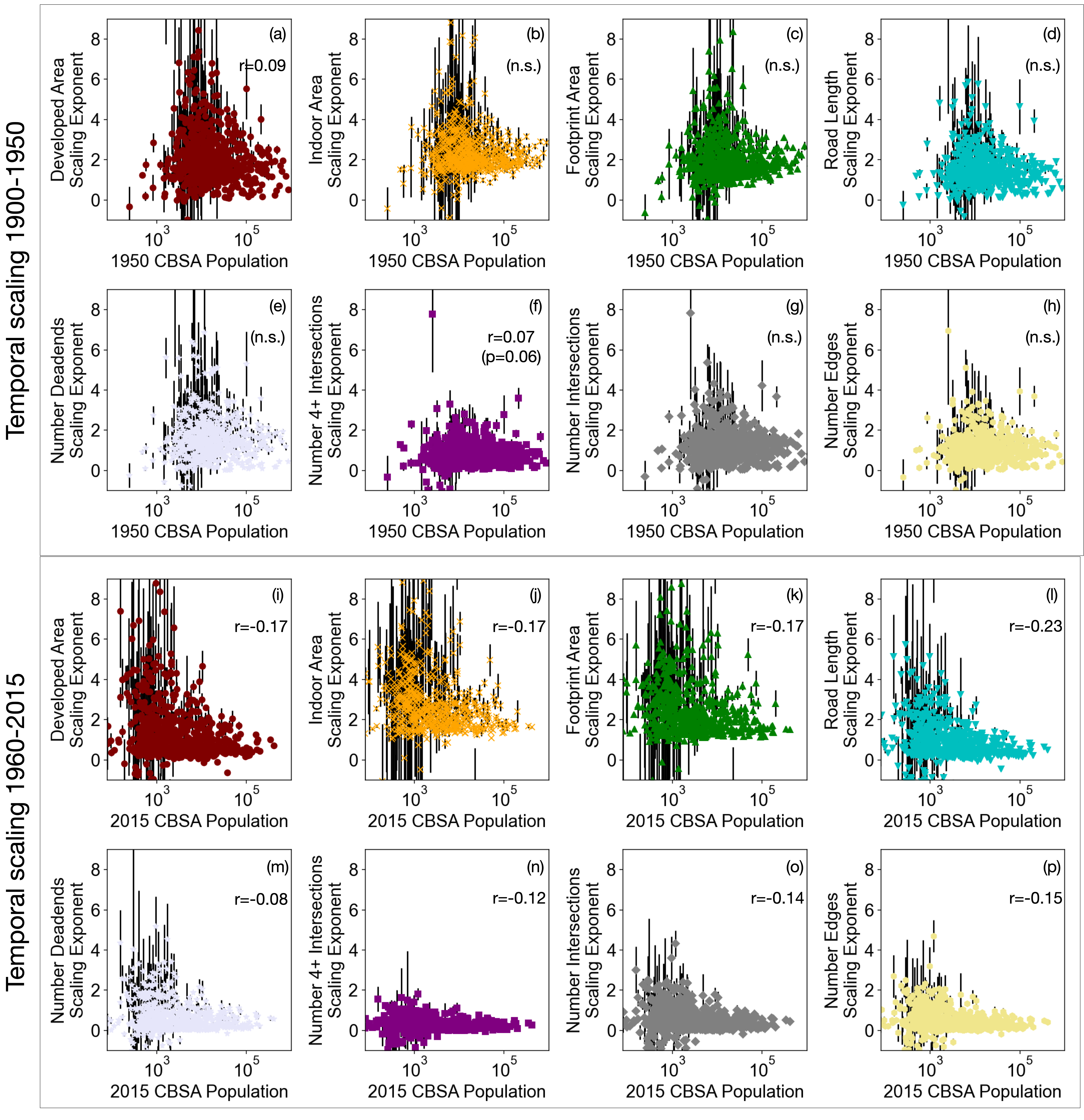}
    \caption{Relationship between scaling exponents and population in 1950 and 2015. (Top panel) scaling law exponent for temporal scaling 1900-1950 versus 1950 population and (bottom panel) scaling law exponent for temporal scaling 1960-2015 versus 2015 population for MSAs with temporal completeness $>60\%$ and geospatial coverage $>40\%$ as in the main text. Top panel: (a,i) Developed area, (b,j) indoor area, (c,k) building footprint area, (d,l) road length, (e,m) number of deadends, (f,n) number of intersections with four or more edges meeting, (g,o) total number of intersections, and (h,p) number of edges. Black error bars represent standard errors of scaling law coefficients. Statistically significant Spearman correlations (p-values$<0.05$) are shown in each figure, and are otherwise denoted ``n.s.'' (not significant).}
    \label{fig:scalingvspop_1900-2015}
\end{figure*}

\begin{figure*}
    \centering
    \includegraphics[width=\linewidth]{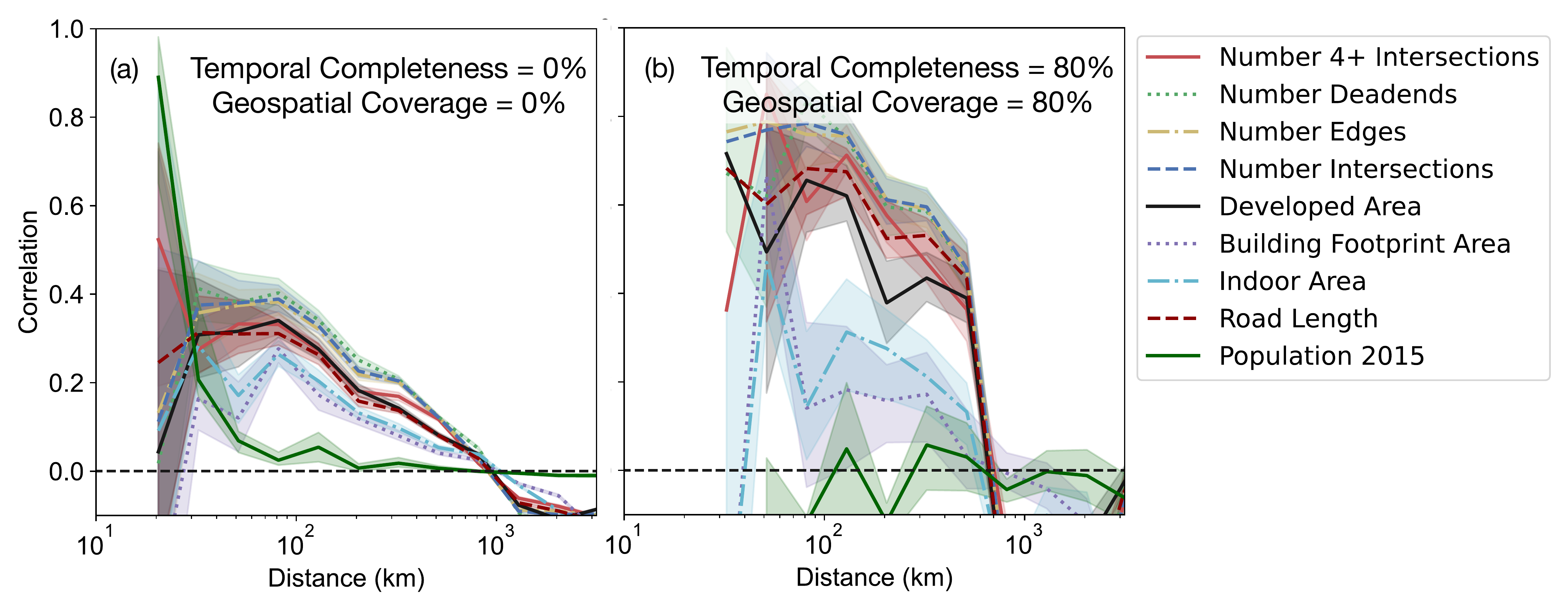}
    \caption{Nearby cities scale similarly.  Figures are the same as main text Fig. 6 except CBSAs have temporal completeness and spatial coverage greater than (a) 0\% and (b) 80\%. Spearman correlation versus distance for the number of 4+ intersections, number of deadends, number of edges, number of intersections, developed area, building footprint area, indoor area, and road length. Shaded regions represent 68\% confidence intervals in the mean correlation.}
    \label{fig:ScalingCorrel_080}
\end{figure*}

\begin{figure*}
    \centering
    \includegraphics[width=0.6\linewidth]{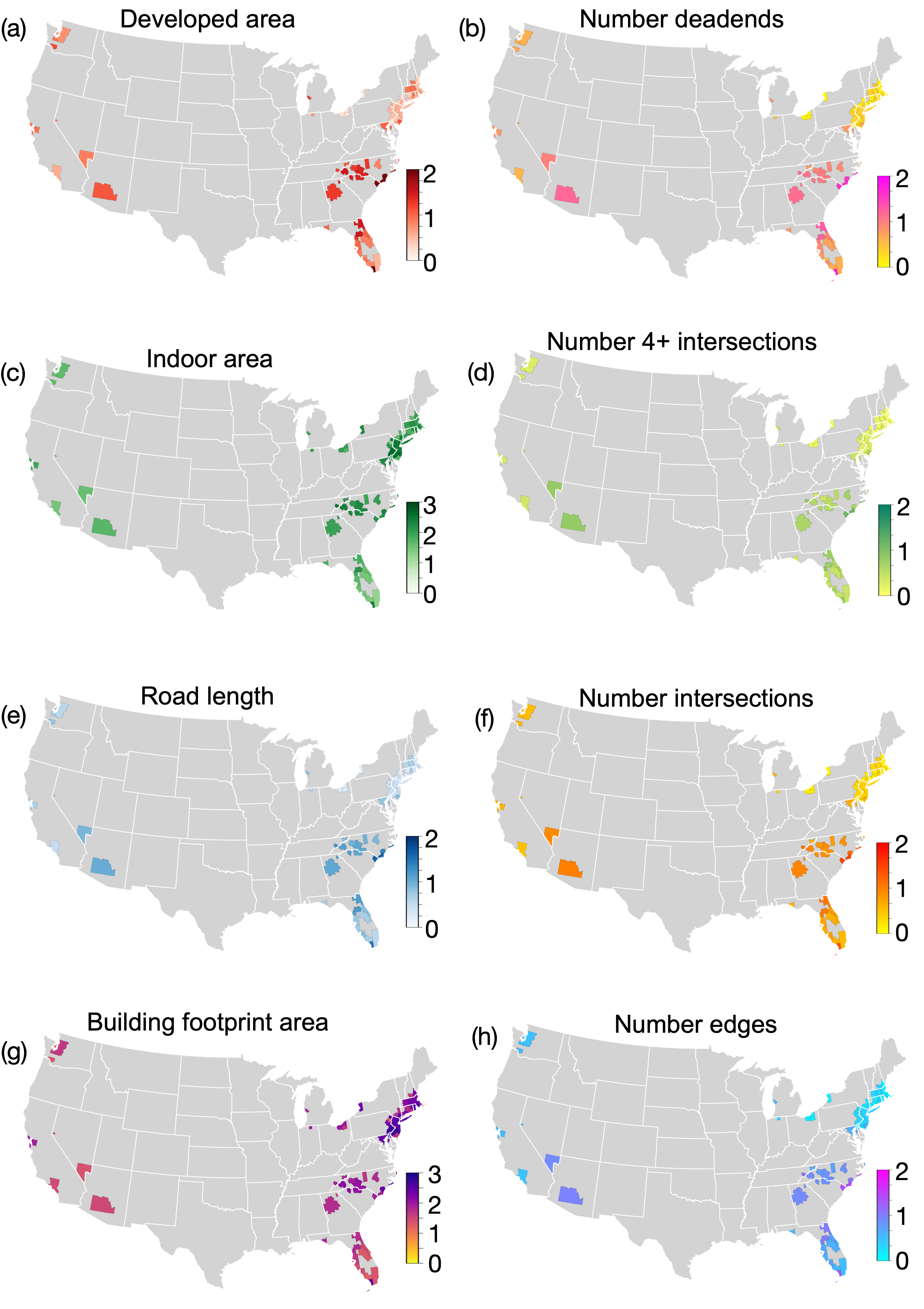}
    \caption{Scaling law exponents across the US for CBSAs with temporal completeness and geospatial coverage $>80\%$. Map of temporal scaling exponents for (a) developed area, (b) number of deadends, (c) building footprint,  (d) number of 4+ intersections, (e) indoor area, (f) total number of intersections, (g) road length,  and (h) number of edges.}
    \label{fig:longitudinal_map_80}
\end{figure*}

\begin{figure*}
    \centering{}
    \includegraphics[width=0.6\linewidth]{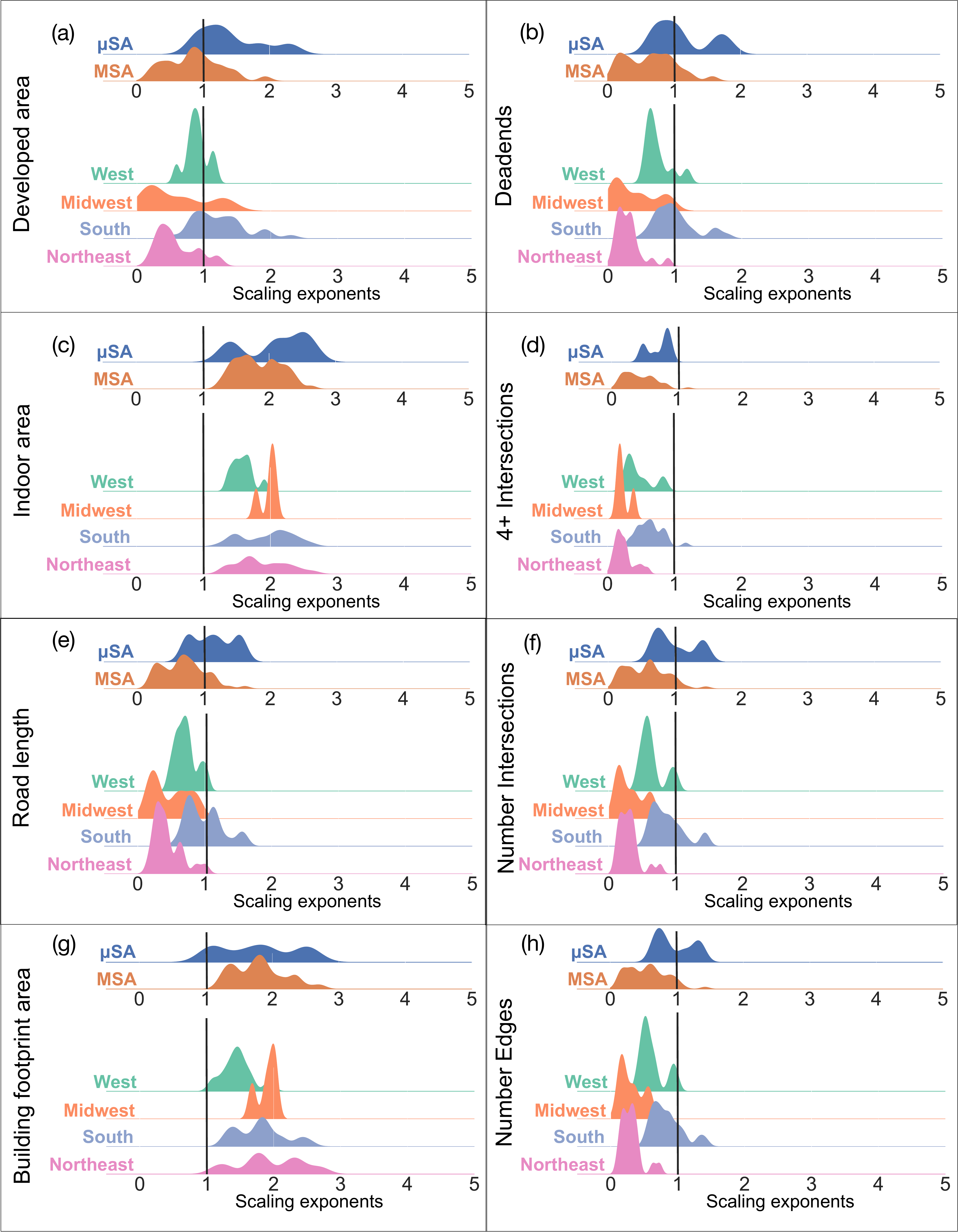}
    \caption{Scaling law exponents split by city size and region for CBSAs with temporal completeness and geospatial coverage $>80\%$. Distribution of temporal scaling exponents for (a) developed area, (b) number of deadends, (c) building footprint,  (d) number of 4+ intersections, (e) indoor area, (f) total number of intersections, (g) road length,  and (h) number of edges.
    Distributions are split by major (MSA) and minor ($\mu$SA) cities as well as OMB-defined regions. Black vertical lines represent linear scaling. }
    \label{fig:Dist_8080}
\end{figure*}

\begin{figure*}
    \centering
    \includegraphics[width=0.6\linewidth]{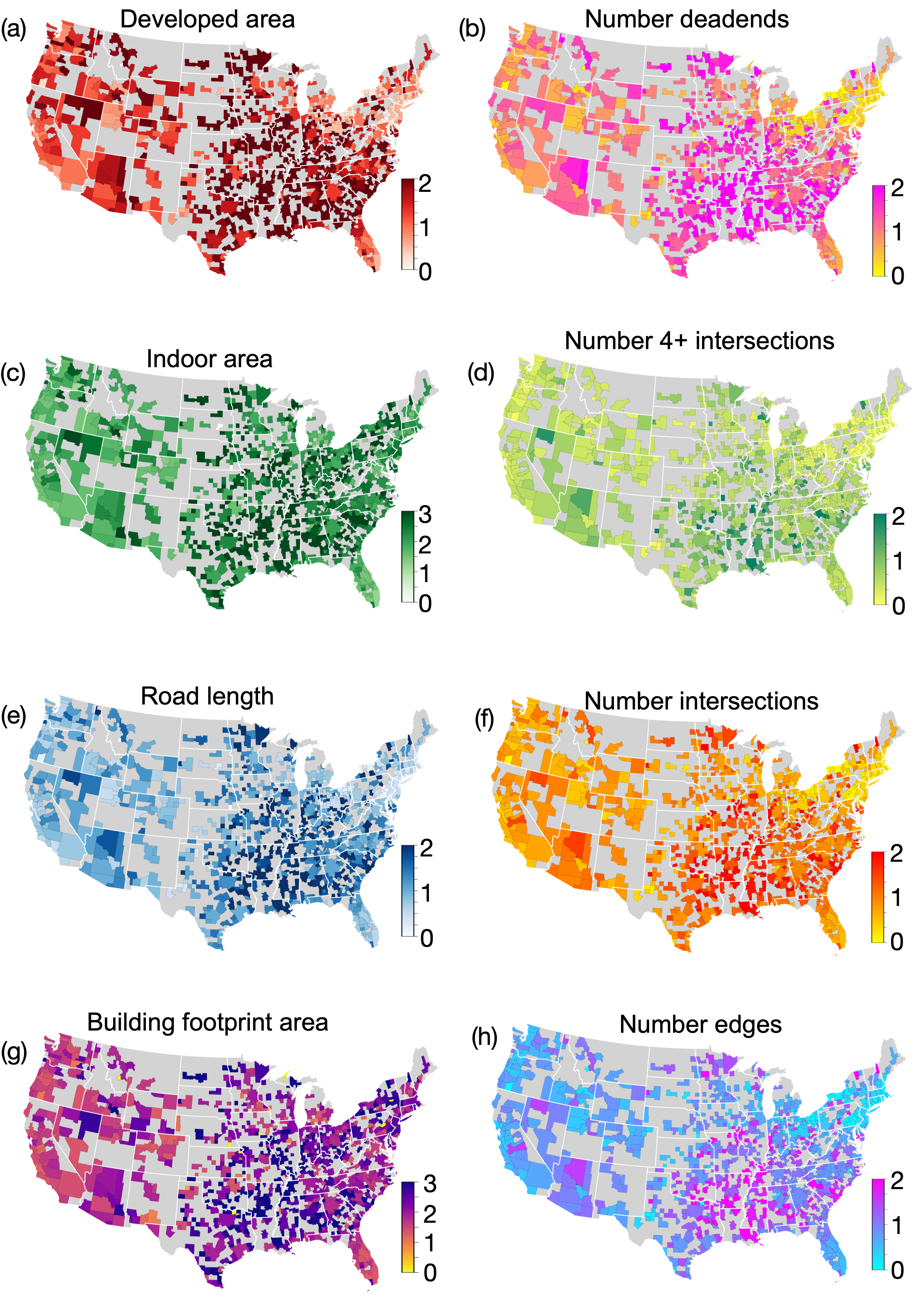}
    \caption{Scaling law exponents across the US for CBSAs with temporal completeness and geospatial coverage $>0\%$. Map of temporal scaling exponents for (a) developed area, (b) number of deadends, (c) building footprint,  (d) number of 4+ intersections, (e) indoor area, (f) total number of intersections, (g) road length,  and (h) number of edges.}
    \label{fig:longitudinal_map_0}
\end{figure*}

\begin{figure*}
    \centering
    \includegraphics[width=0.6\linewidth]{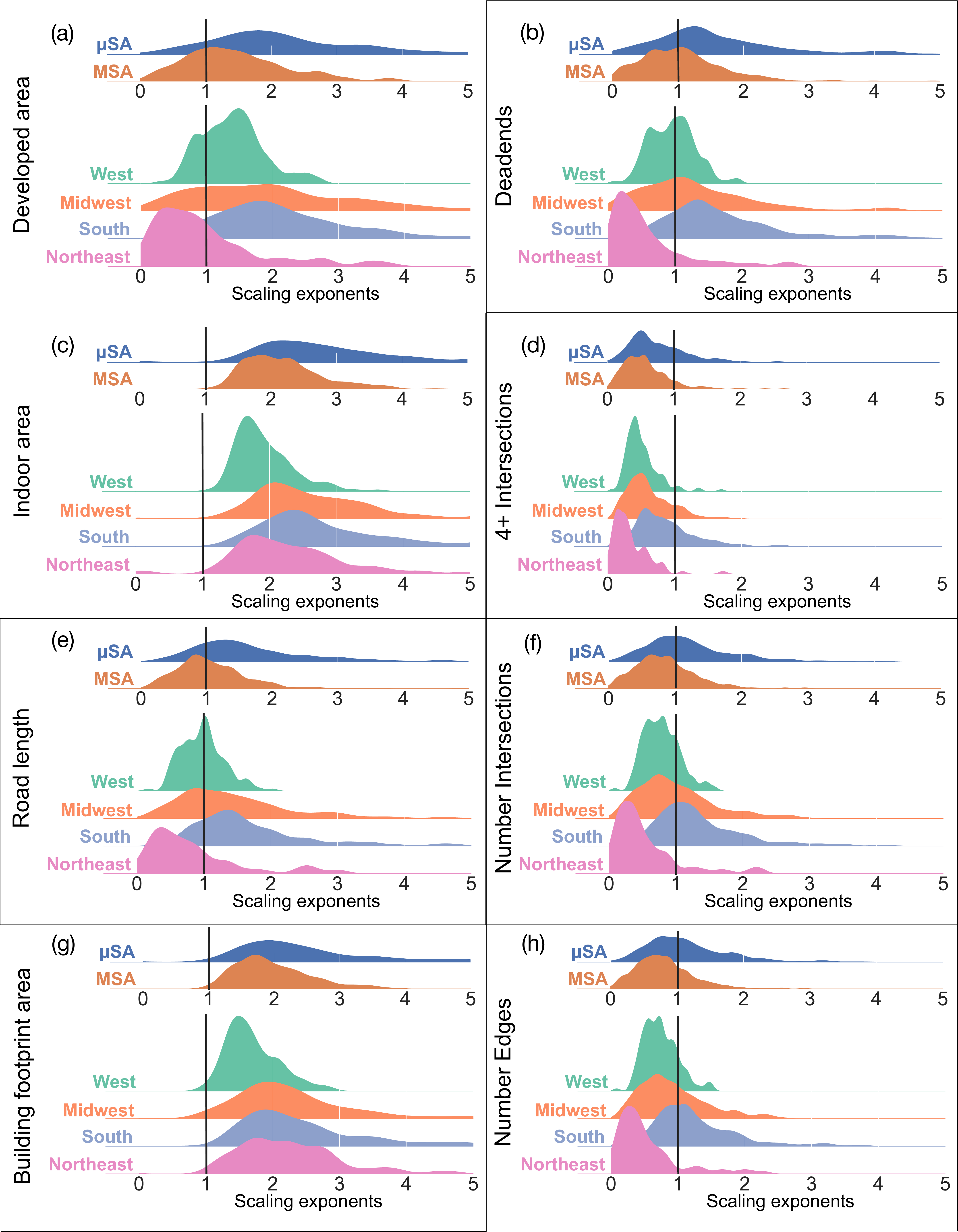}
    \caption{Scaling laws split by city size and region for CBSAs with temporal completeness and geospatial coverage $>0\%$. Distribution of temporal scaling laws for (a) developed area, (b) number of deadends, (c) building footprint,  (d) number of 4+ intersections, (e) indoor area, (f) total number of intersections, (g) road length,  and (h) number of edges.
    Distributions are split by major (MSA) and minor ($\mu$SA) cities as well as OMB-defined regions. Black vertical lines represent linear scaling.}
    \label{fig:Dist_00}
\end{figure*}

\begin{figure*}
    \centering
    \includegraphics[width=0.55\linewidth]{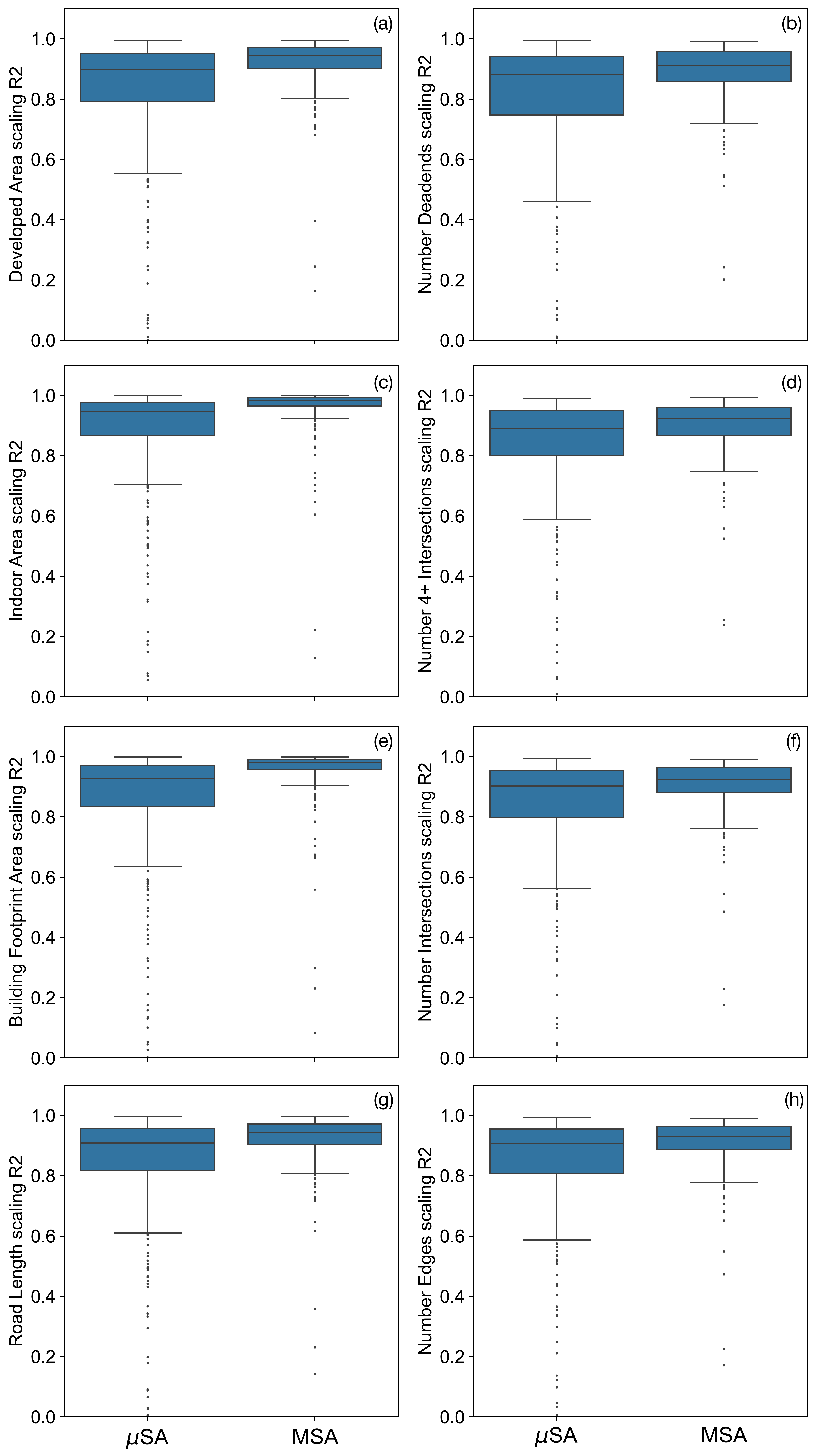}
    \caption{Agreement of temporal relations with power law for cities split by CBSA type (MSA or $\mu$SA) whose temporal completeness is greater than 60\% and spatial coverage is greater than 40\%, as in the main text.  Powerlaw fit $R^2$ for (a) developed area, (b) building footprint, (c) indoor area, (d) road length, (e) number of deadends, (f) number of intersections where four or more edges meet, (g) total number of intersections, and (h) number of edges. Values closer to one correspond to better agreement with power law scaling relation. We find both MSAs and $\mu$SAs have strong fits to the scaling relation, but MSA fits are typically better (Mann-Whitney U test p-value $<0.001$ for each statistic).}
    \label{fig:R2_6040_SA}
\end{figure*}

\begin{figure*}
    \centering
    \includegraphics[width=0.55\linewidth]{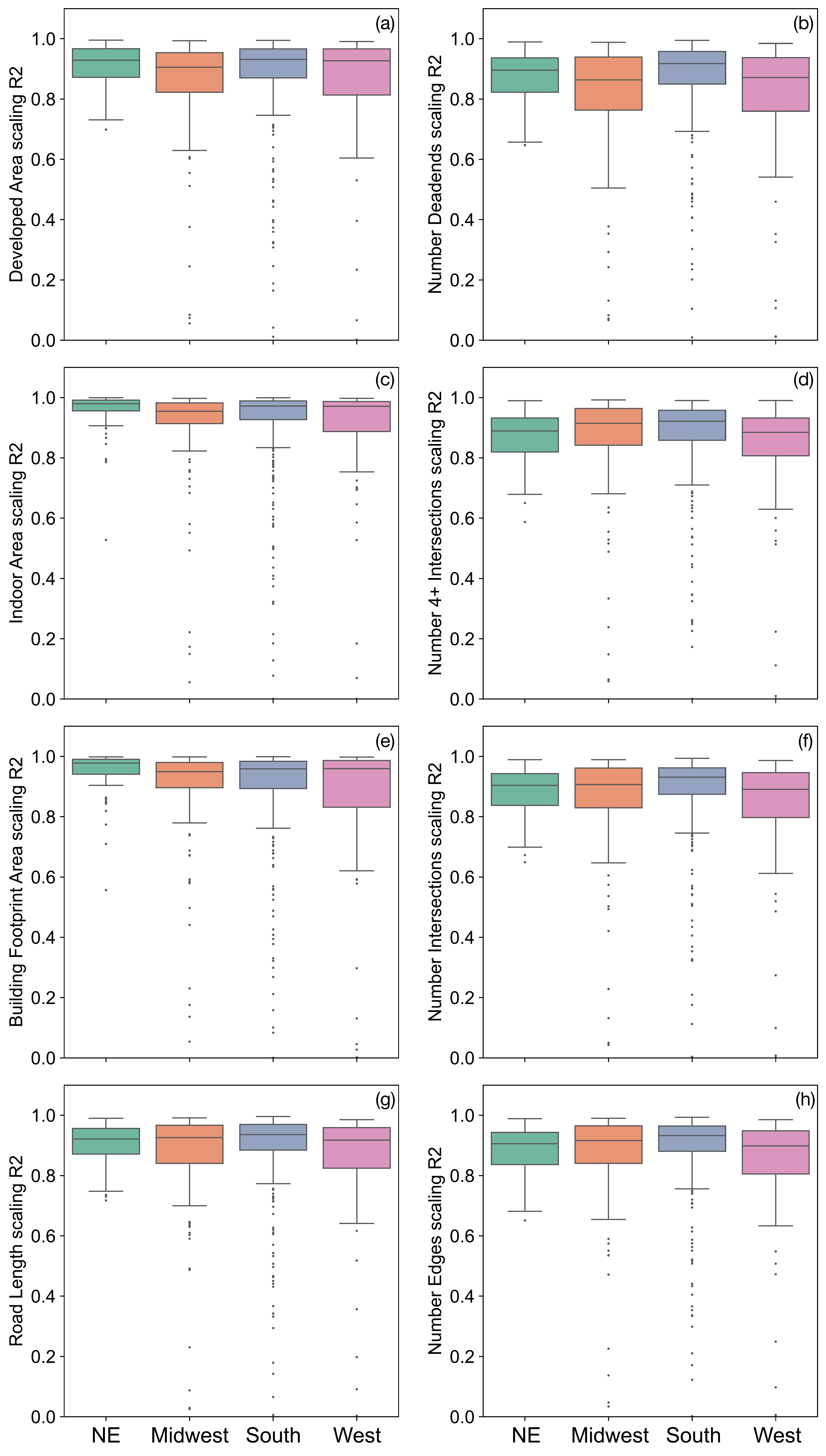}
    \caption{Agreement of temporal relations with power law for cities split by OMB-defined regions whose temporal completeness is greater than 60\% and spatial coverage is greater than 40\%, as in the main text.  Powerlaw fit $R^2$ for (a) developed area, (b) building footprint, (c) indoor area, (d) road length, (e) number of deadends, (f) number of intersections where four or more edges meet, (g) total number of intersections, and (h) number of edges. Values closer to one correspond to better agreement with power law scaling relation. We find strong fits to the scaling relation across regions.}
    \label{fig:R2_6040_region}
\end{figure*}

\begin{figure*}
    \centering
    \includegraphics[width=0.45\linewidth]{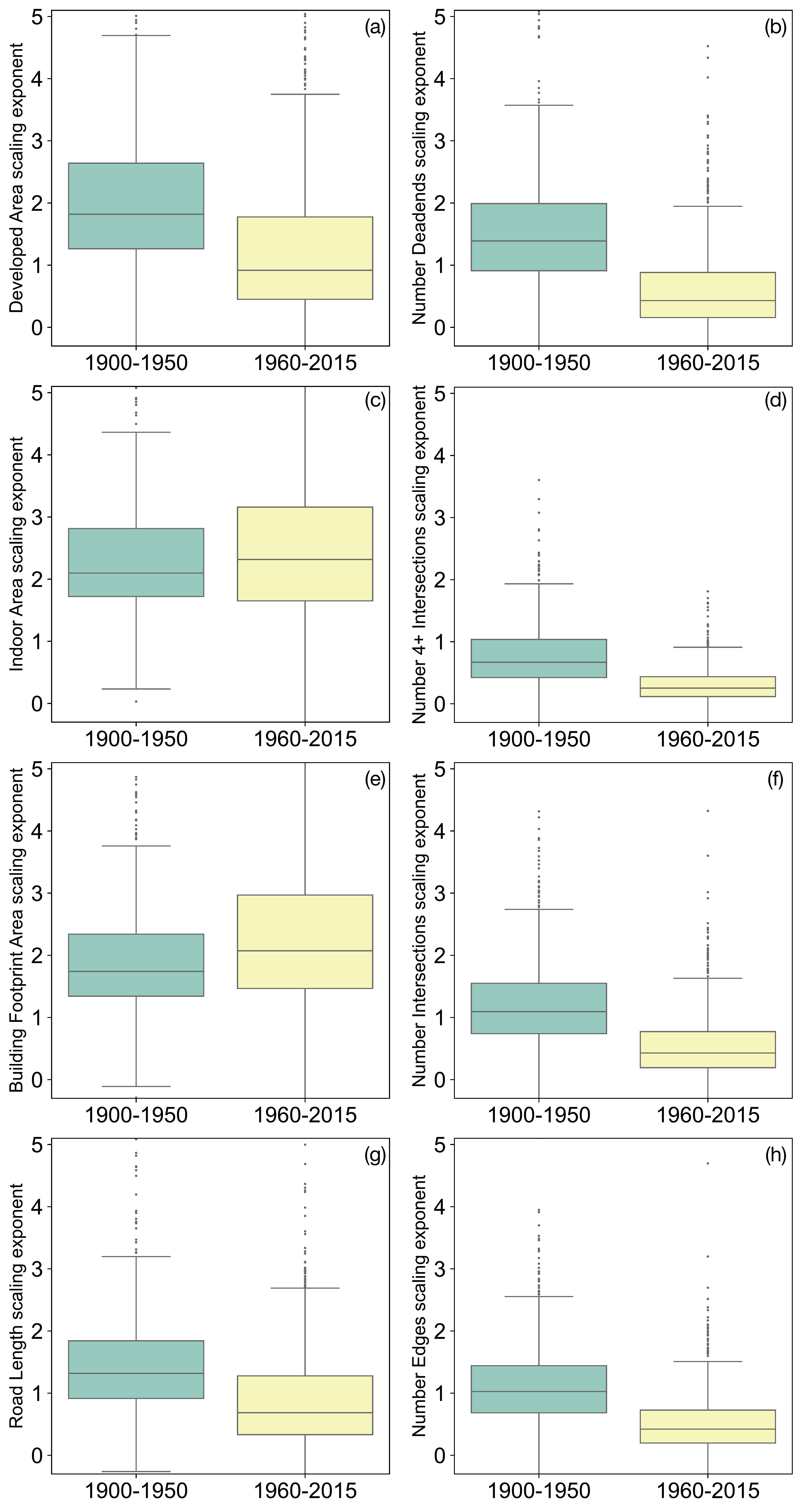}
 \caption{Change in scaling exponents from 1900--1950 and from 1960--2015. Scaling exponents for (a) developed area, (b) building footprint area, (c) indoor area, (d) road length, (e) number of deadends, (f) number of intersections where four or more edges meet, (g) total number of intersections, and (h) number of edges. In agreement with main text Fig. 4, we see that scaling laws are higher for MSAs (Mann-Whitney U test p-values $<0.001$ except for indoor area scaling exponents, p-value$=0.2$). In this analysis, we filter cities to have geographic coverage greater than 40\% and temporal completeness greater than 60\% as in the main text.}
    \label{fig:scaling_change}
\end{figure*}

\end{document}